\documentclass[12pt]{iopart}
\usepackage{graphicx, subfigure, setstack, cite,xcolor,soul}

\usepackage{iopams}  
\pdfoutput = 1
\begin{document}

\title[Finite-size scaling in 2D incompressible active fluids]{Finite-size scaling and double-crossover critical behavior in two-dimensional incompressible polar active fluids}

\author{Wanming Qi$^{1,2,3}$, Lei-Han Tang$^{3,4}$ and Hugues Chat{\'e}$^{3,5}$}

\address{$^1$ Department of Physics, Chung-Ang University, Seoul 06974, Korea}
\address{$^2$ School of Materials Science and Physics, China University of Mining and Technology, Xuzhou, Jiangsu 221116, China}
\address{$^3$ Complex Systems Division, Beijing Computational Science Research Center, Beijing 100193, China}
\address{$^4$ Department of Physics, Hong Kong Baptist University, Hong Kong SAR, China}
\address{$^5$ Service de Physique de l'Etat Condens{\'e}, CEA, CNRS, Universit{\'e} Paris-Saclay, CEA-Saclay, 91191 Gif-sur-Yvette, France}

\ead{\mailto{qiwanming@cau.ac.kr}, \mailto{lhtang@hkbu.edu.hk} and \mailto{hugues.chate@cea.fr}}

\begin{abstract}
We study the order-disorder transition in two-dimensional incompressible systems of motile particles with alignment interactions through extensive numerical simulations of the incompressible Toner-Tu (ITT) field theory and a detailed finite-size scaling (FSS) analysis. The transition looks continuous in the explored parameter space, but the effective susceptibility exponent $\gamma/\nu$ and the dynamic exponent $z$ exhibit a strong, non-monotonic variation on the system size in the form of double crossovers.
At small sizes, mean-field exponents are observed for the homogeneous $k=0$ mode whereas spatial fluctuations follow Gaussian statistics. A first crossover marks the departure from this regime to 
one where the system behaves like the equilibrium XY model with long-ranged dipolar interaction and vortex excitations.
At larger sizes, scaling deviates from the dipolar XY behavior and a second crossover is observed, 
to presumably the asymptotic ITT universality class. At this crossover to genuinely off-equilibrium behavior, advection comes in to expedite transport of fluctuations, suppress large-scale fluctuations and help stabilize long-range order. We obtain estimates and bounds of the universal Binder cumulant and exponents of the ITT class. We propose a reduced hydrodynamic theory, previously overlooked, that quantitatively describes the first scaling regime.
%,{\color{red} with exact finite-size scaling exponent $\beta/\nu = d/4$ and dynamic exponent $z=d/2$ at criticality in $d$ dimensions.}
%%%HC not sure the sentence below is useful here
%numerics further reveal how statistics deviate from the prediction of this theory near the first crossover, which calls for theoretical explanation. 
%We also demonstrate that the conventional theoretical approach to finite-size crossovers, namely the renormalization group flow equations, needs to be interpreted carefully in light of 
%shows limitations in quantitatively capturing 
%the observed behavior. 
By providing a relatively comprehensive numerical picture and a novel analytical description, our results help elucidate finite-size effects in critical active matter systems, which have been argued to be relevant for understanding scale-free behavior in real flocks or swarms. 
\end{abstract}

%
% Uncomment for keywords
\vspace{2pc}
\noindent{\it Keywords\/}: active matter, phase transitions, incompressible fluids, finite-size scaling, crossover phenomena
%
% Uncomment for Submitted to journal title message
\submitto{\NJP}
%
% Uncomment if a separate title page is required
\maketitle
% 
% For two-column output uncomment the next line and choose [10pt] rather than [12pt] in the \documentclass declaration
%\ioptwocol
%

\section{Introduction \label{sec:intro}}

Active matter systems such as bird flocks, insect swarms, sheep herds and cell sheets are driven out of equilibrium by  energy injection at the single-particle level. It is of fundamental interest to understand the emergent collective behavior of such interacting non-equilibrium systems \cite{ramaswamy2010mechanics,marchetti2013,LesHouchesOUP}. One of the simplest, but most striking, emergent phenomenon of active matter is flocking, where many self-propelled 
individuals interacting via short-range alignment interactions\cite{TTR2005,cavagna2017,chate2020dry}
self-organize into collective motion with long-range orientational polar order. Flocking in two dimensions (2D) is an especially fascinating non-equilibrium phenomenon that cannot happen at equilibrium, because the spontaneous breaking of a continuous symmetry is then forbidden by the Mermin-Wagner theorem \cite{MW1966}. 

The past decades have seen intense investigation on the non-equilibrium phase transition from random motion to flocking, based on particle-level systems such as the Vicsek model \cite{vicsek1995} and the associated coarse-grained continuum description, the Toner-Tu field theory \cite{TT1995,TUT1998,TT1998,TTR2005,toner2012}. Most of this work was done in 2D, and demonstrated 
that the order-disorder scenario is in fact best described as a phase-separation phenomenon 
between a disordered gas and an 
orientationally-ordered liquid, with a coexistence phase that makes the transition appear discontinuous at finite-size
 \cite{GC2004,solon2015phase,ginelli2016,chate2020dry,martin2021}.
The self-propelled motion of active particles causes a dynamical rewiring of the interaction network, in contrast to the frozen network of a passive system on lattice; the interplay between alignment interactions and activity-induced rewiring gives rise to density-velocity couplings that make the ordered liquid unstable at onset
 \cite{BDG2009,MBM2010,martin2021}. This prevalence of discontinuous transitions at finite-size in models 
 and theories is in contrast with the advocated criticality of natural swarms \cite{cavagna2017}.

The incompressible Toner-Tu (ITT) field theory \cite{CTL2015,CLT2016,CLT2018,CLT2018squeezed} can be seen as an extreme case of vanishing density fluctuations. Indeed, it does not involve the density field anymore, being a stochastic partial differential equation
governing a velocity field {\bf v}:
\begin{eqnarray}
\partial_t {\bf v} + \lambda ({\bf{v}} \cdot \nabla) {\bf{v}} = -\nabla{\cal P} - (a + b |{\bf{v}}|^2) {\bf{v}} 
 + \mu \nabla^2 {\bf{v}} 
 + {\bf f}({\bf{r}}, t) \label{eqn:tt0}
\end{eqnarray}
where the pressure ${\cal P}$ enforces the incompressibility condition $\nabla\cdot{\bf v}=0$,
${\bf f}$ is a Gaussian white noise, $\lambda$, $\mu$, $b$ are positive constants and the sign of $a$ governs the onset of order at mean-field (deterministic) level.
Incompressibility is a crucial simplifying assumption that has been argued to be relevant in various real situations when density fluctuations are suppressed by repulsion or intelligence (see \cite{CLT2018} for more discussion). Recently some authors have further argued that incompressibility may be a valid approximation when a limited system size prevents the development of inhomogeneities near the onset of collective motion \cite{CCGGP2021}.
The one-loop dynamical renormalization group (RG) analysis of the ITT model near 4D showed that the order-disorder transition is then continuous, and belongs to a distinct non-equilibrium universality class 
% with the critical exponents estimated to order $\epsilon \equiv 4-d$ 
 \cite{CTL2015}. 

The ITT transition universality class as found in \cite{CTL2015} has the interesting feature that both the advective nonlinearity and the ferromagnetic nonlinearity are relevant. Furthermore, the RG analysis near 4D suggested two other known universality classes associated with hyperbolic fixed points on the critical surface, namely the fixed point of a randomly stirred fluid and that of an isotropic ferromagnet with long-ranged dipolar interactions. %(\fref{fig:intro}).
In the case of weak advection on small scales as compared to alignment, the ITT model at finite-size criticality involves an equilibrium to off-equilibrium crossover, following the RG flow from the 
%unstable 
ferromagnetic fixed point to the stable ITT fixed point. This crossover has been argued to be relevant in 3D to the scale-free behavior of animal groups: The near-ordering phase of the Vicsek model at intermediate system sizes can describe certain collective behavior of insect swarms \cite{ACCGMPPRSSV2014}, and the Vicsek model at its finite-size pseudo-critical point was argued to display different dynamic scaling behaviors in the low-activity and high-activity regimes, with the dynamic exponent $z$ in agreement with the ferromagnetic and ITT fixed points respectively \cite{CCGGP2021}. 

The order-disorder transition of the ITT field theory in 2D remains unexplored. 
In particular, it remains unclear whether the discontinuous transition of the 2D compressible Toner-Tu model really turns into a continuous one upon imposing the incompressibility constraint, as the one-loop RG predictions for the ITT universality class are generally not expected to be reliable in 2D. Similarly, it is not clear whether the equilibrium to off-equilibrium crossover argued to be relevant for 3D natural swarms is present in 2D.

\begin{figure}
\centering
\includegraphics[width=12 cm]{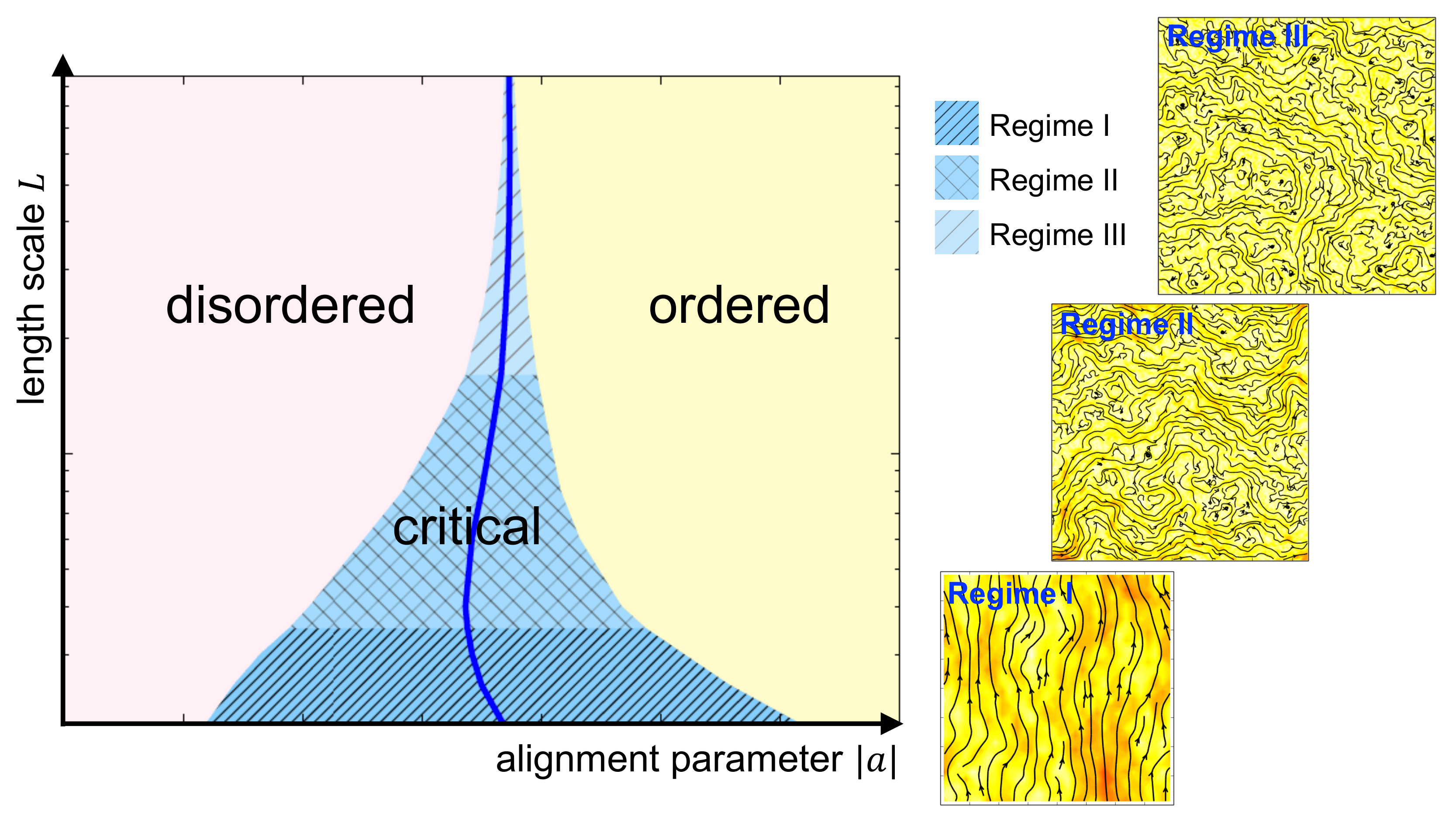}
\caption {Schematic finite-size phase diagram for the two-dimensional incompressible Toner-Tu (ITT) model. The system undergoes a phase transition to long-range orientational order, as alignment interactions strengthen. This transition looks continuous in our numerical simulations for all the explored system sizes $L$. The solid line indicates location of susceptibility peak at a given $L$, flanked by a critical region of size
$\Delta a\simeq L^{-1/\nu}$ (see equation (\ref{eq:delta-a})). The critical phenomena at different lengthscales fall into one of the three regimes: Regime I is in the vicinity of the Gaussian fixed point, and can be described by a kind of reduced hydrodynamics as constructed in this paper; Regime II is characterized by the equilibrium universality class of dipolar XY, and Regime III corresponds to the non-equilibrium class of ITT. Typical velocity field patterns of different regimes are visualized with streamlines (flow speed in color); Regime I features the vortex-free patterns in contrast to the other two. }  
\label{fig:intro}
\end{figure}

In this work, we focus on the emergence of flocking in the 2D ITT field theory.
We numerically integrate equation~(\ref{eqn:tt0}) to study the order-disorder transition, focusing attention on the finite-size scaling (FSS) of the steady-state statistics of both the global speed (termed mean flow) and the spatially-varying inhomogeneous modes. The phase transition does appear continuous in the explored parameter space. As shown in \fref{fig:intro}, we observe three distinct scaling regimes near criticality as the linear system size $L$ increases. They are denoted as ``Regime I", ``Regime II" and ``Regime III" in increasing order of size. In Regimes I and II, the ITT model behaves as the continuum model that describes at steady state the long-distance behavior of the equilibrium dipolar XY system (the latter is simulated here as ITT with advection turned off), suggesting that advection is negligible for these smaller systems. Regime I is further marked by the absence of vortex excitations as seen in \fref{fig:intro}. By comparing with Monte Carlo simulation of dipolar XY \cite{MC2010}, we identify Regime II as the scaling regime of the dipolar XY class that contains vortex-like excitations in the flow field. In Regime III, ITT deviates from the dipolar XY behavior, as advection comes in to cause faster transport of fluctuations across the system, suppress large-scale fluctuations and help stabilize collective motion. Regime III corresponds to the asymptotic, non-equilibrium universality class of 2D ITT, and we obtain estimates and bounds for its Binder cumulant and some critical exponents. The crossover behavior indicates that the dipolar ferromagnet fixed point is a hyperbolic fixed point on the critical surface for 2D, which was also found near 4D \cite{CTL2015}.  

In Regime II or Regime III, as is conventionally expected, the homogeneous $k = 0$ mode and the inhomogeneous $k > 0$ modes show the same set of universal exponents that characterizes a specific RG fixed point. However, this is not the case for Regime I with a visible mean flow (figure 1). This small-size equilibrium regime shows the interesting feature that mean-field exponents are observed for the $k = 0$ mode, whereas the Gaussian statistics is found for $k > 0$.  Inspired by the numerical observation of the velocity field patterns near criticality, we propose a reduced hydrodynamic theory which we term the ``weak-coupling critical theory" (WCCT). In the WCCT, the eddies (i.e., the $k > 0$ modes) are statistically described by the critical linear theory (or, the Gaussian fixed point), and the effective dynamics of the mean flow (i.e., the $k = 0$ mode) is derived by averaging out the eddies \footnote{Following the convention in fluid mechanics, the homogeneous $k = 0$ mode is called the ``mean flow", and the inhomogeneous $k > 0$ modes of the incompressible fluid are the ``eddies".}. Since the advective nonlinearity respects Galilean invariance as in Navier-Stokes hydrodynamics, it is completely neglected in the WCCT. The ferromagnetic nonlinearity violates Galilean invariance and weakly couples the eddies to the mean flow, which results in a size-dependent shift of the critical point. Despite this %striking difference of 
a shifting critical point which needs to be treated with care,
%as opposed to a constant infinite-size critical point, 
the WCCT leads to exact FSS forms for the $k = 0$ statistics similar to the conventional large-size FSS forms, with mean-field exponents $\gamma = 1$, $\nu = 1$, $\beta = 1/2$, $z = 1$; the Binder cumulant at the shifting critical point $G_{\rm c} = 1-\pi/6 \approx 0.4764$. The predictions agree well with the simulations for the smallest systems in Regime I. Interestingly, the dynamic exponent $z =1$ of our WCCT agrees
with $z = d/2$ at the ``non-dissipative'' fixed point by Cavagna {\it et al}\cite{cavagna2019,cavagna2019PRE} in 2D, and the latter in 3D was suggested to be most relevant for the observed collective behavior of natural swarms.

%$z = d/2$ of our WCCT agrees with those at the ``non-dissipative'' fixed point by Cavagna {\it et al}\cite{cavagna2019,cavagna2019PRE}, which was suggested to be most relevant for the observed collective behavior of natural swarms. 
%%%HC remark below best placed in dsicussion section
%Future studies may clarify this possible correspondence.

%Within the RG framework, scaling properties in each regime and the crossover between different scaling regimes are described by exponents characterizing the relevant unstable directions of the RG flow\cite{nattermann1992}. While our numerical results are generally in line with the predictions of the one-loop calculation in \cite{CTL2015}, there are also important differences both at qualitative and quantitative level. 

We have also examined the extent at which the one-loop RG calculation in \cite{CTL2015} captures scaling properties in the three regimes described above, as well as the crossovers. Broadly speaking, given its perturbative nature in the strength of the nonlinear terms in equation (\ref{eqn:tt0}), predictions of the one-loop RG are not expected to be quantitative unless one is close to the upper critical dimension or in the neighborhood of the Gaussian fixed point\cite{msr1973,nattermann1992}.
Our results support this view in general but also bring out an interesting and subtle feature of the ITT model, i.e., the $k = 0$ and $k>0$ modes have different scaling properties in Regime I. Only the latter are approximately described by the RG flow near the Gaussian fixed point.

%The next question is whether RG can capture the crossover from Regime I to Regime II, by a flow trajectory starting near the Gaussian fixed point to approach the dipolar XY fixed point. Although the perturbative RG analysis near 4D \cite{CTL2015} is generally not expected to be accurate in 2D, we rationalize our quantitative assessment by noting that the one-loop flow equations may be accurate enough near the Gaussian fixed point when nonlinearities are weak. {\color{blue}We find that the one-loop RG trajectory agrees with numerical simulation %at best 
%qualitatively, but fails to quantitatively reproduce the value of the crossover size.} 
%%%HC the following best said in conclusion/discussion
%Besides the crossover size, numerical simulation provides more information on how statistics deviate from the WCCT prediction and depend on the growing ferromagnetic nonlinearity; we leave to future theoretical investigation to explain these observations.    

The remainder of this paper is organized as follows. \Sref{sec:dns} details our numerical methods. The equation of motion (EOM) is given in two representations: the conventional one in terms of the velocity field which is used in numerical simulations, while the other one, in terms of vorticity and streamfunction, facilitates analytical calculations. \Sref{sec:crossover} presents the numerical observations of the double-crossover critical behavior in both static and dynamic scaling. At the end of this section, our measured exponents are compared with the one-loop RG predictions of \cite{CTL2015}, and the equilibrium to off-equilibrium crossover is further discussed, with a scaling argument and a summary of the non-equilibrium effects of advection. In \sref{sec:mft} we address the problem of theoretically describing finite-size effects in small systems, by constructing the WCCT and comparing it against numerical simulation. Some discussion and open questions are presented in \sref{sec:conclusions}.

\section{Model and Numerical Methods \label{sec:dns}}

Equation~(\ref{eqn:tt0}) can be rewritten:
\begin{eqnarray}
\partial_t {\bf{v}} = \mathbb{P}\{- \lambda ({\bf{v}} \cdot \nabla) {\bf{v}} - (a + b |{\bf{v}}|^2) {\bf{v}} + \mu \nabla^2 {\bf{v}}  + {\bf f}({\bf{r}}, t)\}, \label{eqn:tt}
\end{eqnarray}  
where the projection operator $\mathbb{P} $ enforces incompressiblity by projecting out the compressible Fourier modes:
\begin{eqnarray}
\mathbb {P} {\bf{F}}({\bf{r}}) &\equiv& \int \frac{\rmd^2 {\bf{k}}}{(2 \pi)^2} ~\rme^{\rmi {\bf{k}} \cdot {\bf{r}}} (\mathbb{I} - {\bf{k}}  \frac{{\bf{k}} \cdot}{k^2}) {{\bf{F}}}({\bf{k}}) \;,
\end{eqnarray}
%This projection is equivalent to the pressure term enforcing the incompressibility condition $\nabla \cdot {\bf v} = 0$ in the conventional form of the ITT equation \cite{CTL2015}, and this can be shown by taking the divergence of both sides of \eref{eqn:tt} and solving the pressure in Fourier space. 
and the noise correlations are
\begin{eqnarray}
\langle f_i({\bf{r}}, t) f_j({\bf{r'}}, t')\rangle = 2D \delta_{ij} \delta^2({\bf{r}} - {\bf{r'}}) \delta(t-t'), \label{eq:noisef}
\end{eqnarray}
where the noise strength $D$ is another parameter of the model, and $i, ~j = x, ~y$ denote Cartesian components. 

We perform pseudo-spectral simulations on an $L\times L$ box with doubly periodic boundary conditions. The lattice spacing is $\Delta x$. No dealiasing method is used, as the simulations are stable and well-behaved even in the absence of dealiasing. The Euler method is used to integrate the field forward in time with the time step $\Delta t$.  The vector noise at each lattice point and each time step is given by an amplitude and a phase angle each randomly independently drawn from a uniform distribution on $[0, \sqrt{\frac{D_0}{(\Delta x)^2 ~\Delta t }}]$ and on $[0, 2\pi)$ respectively. Note that the noise is not strictly implemented as Gaussian noise to prevent numerical instability due to rare large numbers. Near a continuous phase transition, as the correlation time of fluctuations becomes much larger than the time step, according to the central limit theorem, the noise effectively obeys the Gaussian statistics (\ref{eq:noisef}) with $D = D_0/12$.

The time-dependent order parameter that characterizes collective motion is the norm of the global velocity,
\begin{eqnarray}
c(t) \equiv | {\bf{c}}(t)| \equiv |\langle {\bf{v}}({\bf{r}}, t)\rangle_{\bf{r}}| ,
\end{eqnarray}
where $\langle \cdots \rangle_{\bf{r}}$ denotes averaging over the whole $L\times L$ box. We numerically measure the low-order steady-state statistics of $c(t)$, namely, the mean $\langle c \rangle$, the susceptibility
\begin{eqnarray}
\chi \equiv L^2 [\langle c^2 \rangle - \langle c\rangle^2], \label{eq:susdef}
\end{eqnarray}
and the Binder cumulant 
\begin{eqnarray}
G \equiv 1 - \frac{\langle c^4 \rangle}{3 \langle c^2 \rangle^2}.
\end{eqnarray}
Here the average $\langle \cdots \rangle$ is the long-time and ensemble average.
The system in the thermodynamic limit near 4D was predicted to undergo a continuous phase transition from a disordered phase where $\langle c \rangle$ approaches $0$, to an ordered phase where it approaches a finite value \cite{CTL2015}. 
The existence of true long-range orientational order in 2D ITT was proven in \cite{CLT2016}.

To cross the possible phase transition, we tune the linear coefficient $a$ as the control parameter while keeping the other model parameters fixed. At mean-field level in the absence of noise, $a<0$ is the ordered phase and $a>0$ is the disordered phase. The parameters $D_0 = \mu = 1$ and $b \approx 0.23$ \footnote{The parameter $b$ is set as $b =  - (\frac{1}{3} + e^{-1/2}) \cdot(\frac{2}{5} - e^{-1/8}) \cdot(\frac{16}{15} + 2  e^{-1/2}- e^{-1/8})^{-2} \approx 0.23$. }. To gain insight into the effect of advection, two values of $\lambda$ are used, namely $\lambda = 5$ for the so-called ``ITT-5" simulations, and $\lambda = 10$ for the ``ITT-10" simulations. 
For comparison, we also perform another set of simulations with the same $\{ D_0, \mu, b\}$ but $\lambda = 0$, which corresponds to the continuous dipolar XY model. These simulations with advection turned off will be called the ``ferro" simulations. Various spatial resolutions $\Delta x$ and time steps $\Delta t$ are used (see table \ref{table:dxdt}).

\fulltable{ \label{table:dxdt}Summary of simulation parameters and resolutions: the advection coefficient $\lambda$, lattice spacing $\Delta x$, time step $\Delta t$, the number of grid points in each dimension $N$, and the range of total integration time $T$. }
\br
Simulation & $\lambda$ & $\Delta x$ &   $\Delta t$  &  N  & T \\
\mr
ferro-s &  0 & 0.5 &   0.01 &    12, 16, 24, 32 & $5 \times 10^5 \sim 9 \times 10^6$\\
ferro-m & 0 & 5.0 &  0.01 & 12, 16, 24, 30, 48 &  $5 \times 10^5 \sim 3 \times 10^7 $ \\
 ferro-l1 & 0 & 25.0&  0.05 &  16, 32, 40  & $2 \times 10^7 \sim 3 \times 10^8 $ \\
 ferro-l2 & 0 & 25.0&  0.2&  64  & $2 \times 10^9 \sim 4 \times 10^9 $\\
ferro-xl & 0 & 50.0 &  0.2 &  16, 20, 32, 48, 64  & $1 \times 10^8 \sim 1 \times 10^{10} $\\
ITT-5s & 5.0 & 0.5 & 0.01 & 12, 20, 32, 48, 64, 96, 128 & $1 \times 10^6 \sim 3 \times 10^7 $\\
ITT-5m & 5.0 & 6.25 & 0.1 &  8, 12, 16, 20, 24, 28, 32, 36,  &   $2 \times 10^6 \sim 3 \times 10^9 $ \\
            &        &         &       & 48, 64, 128, 192, 256, 320 & \\
ITT-10 & 10.0 & 2.5 & 0.05 & 8, 10, 12, 14, 16, 24, 32, 48, & $1 \times 10^6 \sim 4 \times 10^8 $\\
           &         &       &          & 64, 96, 128, 160, 240, 320, 384 & \\
\br
\endfulltable

For analytical calculations, it is more convenient to work in the vorticity-streamfunction formalism. The velocity field is decomposed into the mean flow ${\bf c}(t)$ and the eddies $ {\bf v}' ({\bf r}, t)$. The streamfunction $\psi({\bf r}, t)$ of the eddies is defined by
\begin{eqnarray}
{\bf v} ({\bf r}, t) \equiv {\bf c}(t) + {\bf v}' ({\bf r}, t)\equiv {\bf c}(t) + \hat{\bf z} \times \nabla \psi.
\end{eqnarray}
The doubly periodic boundary condition on ${\bf v}$ translates into the same boundary condition on $\psi$.
Define the scalar vorticity 
\begin{eqnarray}
\omega({\bf r}, t) \equiv \hat{\bf z} \cdot (\nabla \times {\bf v}'), 
\end{eqnarray}
and there is a one-on-one correspondence between $\omega$ and $\psi$ given by 
\begin{eqnarray}
\omega = \nabla^2 \psi \label{eq-omegapsi}
\end{eqnarray}
with the specific boundary conditions.
The EOM in the streamfunction-vorticity representation is given by
 \begin{eqnarray}
 \fl \frac{d {\bf c}}{dt} = (-a - b |{\bf c}|^2 - b ~\langle |\nabla \psi|^2 \rangle_{\bf r}~) {\bf c} - b~ \langle |\nabla \psi|^2 {\bf v}' \rangle_{\bf r}  -  2 b~ \langle  {\bf v}' ({\bf c} \cdot {\bf v}' )\rangle_{\bf r} + \langle {\bf f} \rangle_{\bf r}, \label{eq1}\\
\fl \frac{\partial \omega}{\partial t} = - \lambda ({\bf c} \cdot \nabla) \omega + \lambda J[ \omega, \psi] -a \omega + \mu \nabla^2 \omega + \eta({\bf r}, t) \nonumber\\
\fl ~~~~~~~~ -2 b \hat{\bf z} \cdot [\nabla \times(|\nabla \psi|^2 {\bf c})] - 2 b \hat{\bf z} \cdot \{ \nabla \times [({\bf c} \cdot {\bf v}') {\bf c} ] \} \nonumber\\
\fl ~~~~~~~~  -  b |{\bf c}|^2 \omega - b \hat{\bf z} \cdot [ \nabla \times (|\nabla \psi|^2  {\bf v}'  ) ]  -2 b ({\bf c} \cdot  {\bf v}' ) \omega, \label{eq2}
 \end{eqnarray}
 where the Jacobian 
 \begin{eqnarray}
 J[\omega, \psi] \equiv \frac{\partial \omega}{\partial x} \frac{\partial \psi}{\partial y} -  \frac{\partial \omega}{\partial y}  \frac{\partial \psi}{\partial x},
 \end{eqnarray}
and the vorticity noise $\eta \equiv \hat{\bf z}\cdot ( \nabla \times {\bf  f})$. We have simplified the equations using the identities $|{\bf v}'|^2 = |\hat{\bf z} \times \nabla \psi|^2 = |\nabla \psi|^2$, $\hat{\bf z} \cdot \{ \nabla \times [({\bf c} \cdot {\bf v}') {\bf v}' ]\} = ({\bf c} \cdot {\bf v}')\omega + (1/2) \hat{\bf z} \cdot [\nabla \times (|\nabla \psi|^2 {\bf c})]$, and $\langle {\bf v}' \rangle_{\bf r} = {\bf 0}$. Note that the nonlinear coupling appearing in the EOM of the mean flow is the ferromagnetic nonlinearity, not the advective nonlinearity, because the latter respects Galilean invariance as in Navier-Stokes fluids. The effect of advection on the mean flow is only indirect: Advection affects the dynamics of the eddies, and the ferromagnetic nonlinearity couples the eddies to the mean flow. 

\section{Double-crossover scaling as system size increases \label{sec:crossover}}

\subsection{Numerical observations \label{subsec:DNS}}

\begin{figure}
\centering
\includegraphics[width=14.0 cm]{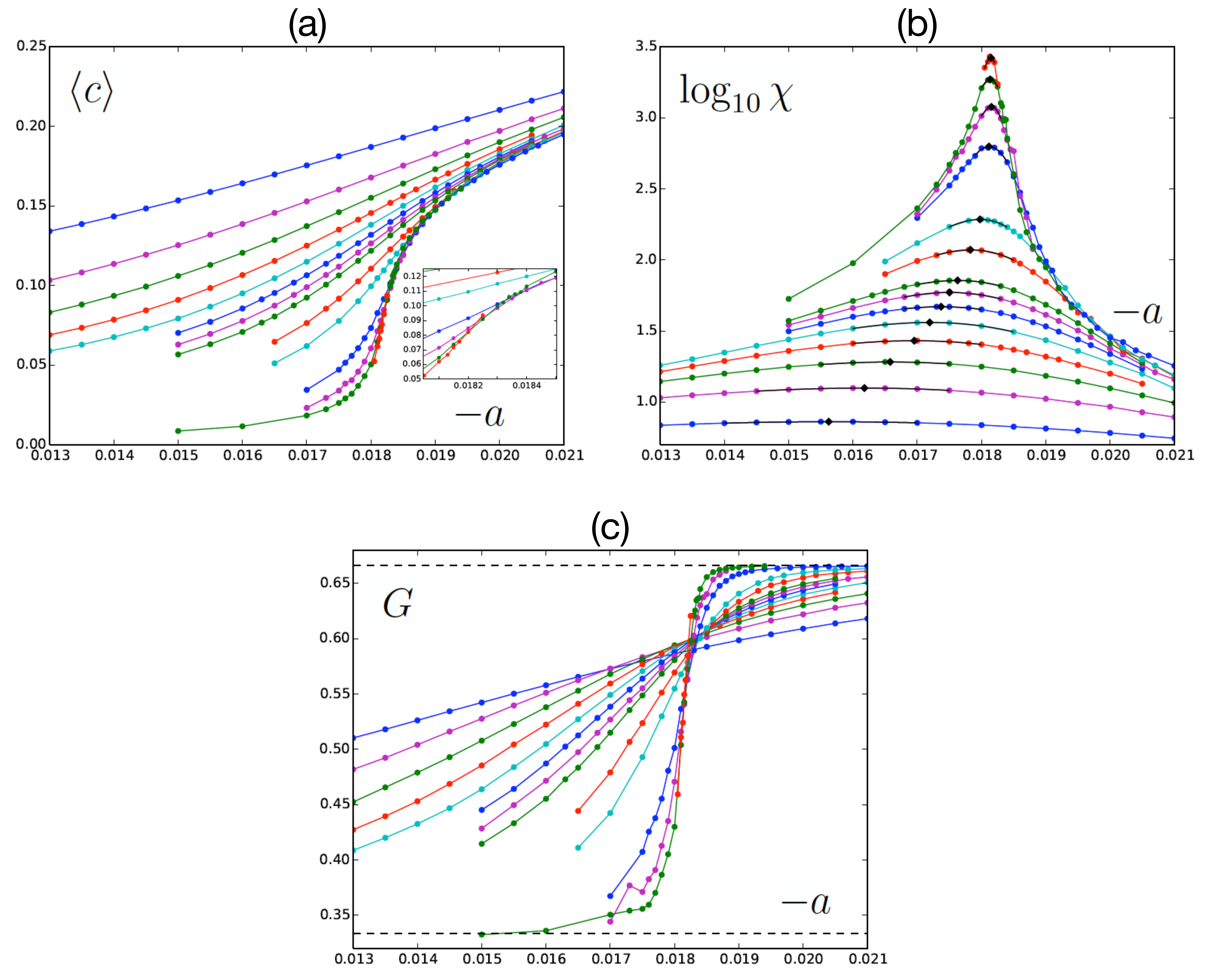}
\caption {(a) Time-and-ensemble-averaged order parameter $\langle c\rangle$, (b) susceptibility $\chi$ in $\log_{10}$ scale, and (c) Binder cumulent $G$ as functions of the control parameter $-a$, for the ``ITT-5m" simulations with different system sizes. Colors cycle from blue, to magenta, to green, to red, to cyan and back to blue, in increasing order of size. That there are no discontinuous jump in $\langle c \rangle$ and no dip in $G$ supports the scenario of a continuous phase transition. In (b) the peak regions are fitted with polynomials shown with black lines, and the peaks are marked as black diamonds. The dashed lines in (c) mark the limiting values $G = 1/3$ for the disordered phase and $G = 2/3$ for the ordered. Inset of (a): a magnified view showing that the largest-$L$ curves cross, which seems a complication due to the finite-size crossover. System sizes: $L = 50, ~75, ~100, ~125, ~150, ~175, ~200, ~225, ~300, ~400, ~800, ~1200, ~1600, ~2000$.}  
\label{fig:trans-five}
\end{figure}

The mean order parameter $\langle c \rangle$, the susceptibility $\chi$ and the Binder cumulant $G$ are plotted against the control parameter $-a$ for the ``ITT-5m" simulations in figure  \ref{fig:trans-five}. When the linear coefficient $-a$ is small, noise dominates over the ferromagnetic alignment interaction, the velocity patterns appear globally random and isotropic, though locally small-scale and short-lived vortices and jets abound. The mean order parameter $\langle c \rangle$ is small in this case. As $-a$ increases, $\langle c \rangle$ increases continuously to order-one values, although the order parameter transition curves cross for the largest four sizes (see figure \ref{fig:trans-five}(a)). As we have checked for several sizes near transition, no bandlike structures of phase separation have been observed in the velocity field snapshots. No bistability has been found in $c(t)$ at the steady state, and the steady-state probability distribution of $c$ is unimodal. This is supported by the behavior of the Binder cumulant $G$ in figure \ref{fig:trans-five}(c): It exhibits monotonic change from the limiting value $1/3$ in the disordered phase to the other limit $2/3$ in the ordered phase, and no dip. Although the crossing of the order parameter curves seems to indicate a first-order transition, here considering the absence of phase separation and the presence of similar curve crossing in susceptibility, we tend to believe that this crossing is due to a finite-size crossover of a continuous transition as will be detailed in the following. Thus these observations all agree with the scenario of a continuous phase transition. The transition curves for the other ITT simulations as well as the auxiliary simulations ``ferro" are shown in \ref{app:trans}.

\begin{figure}
\centering
\includegraphics[width=8.5 cm]{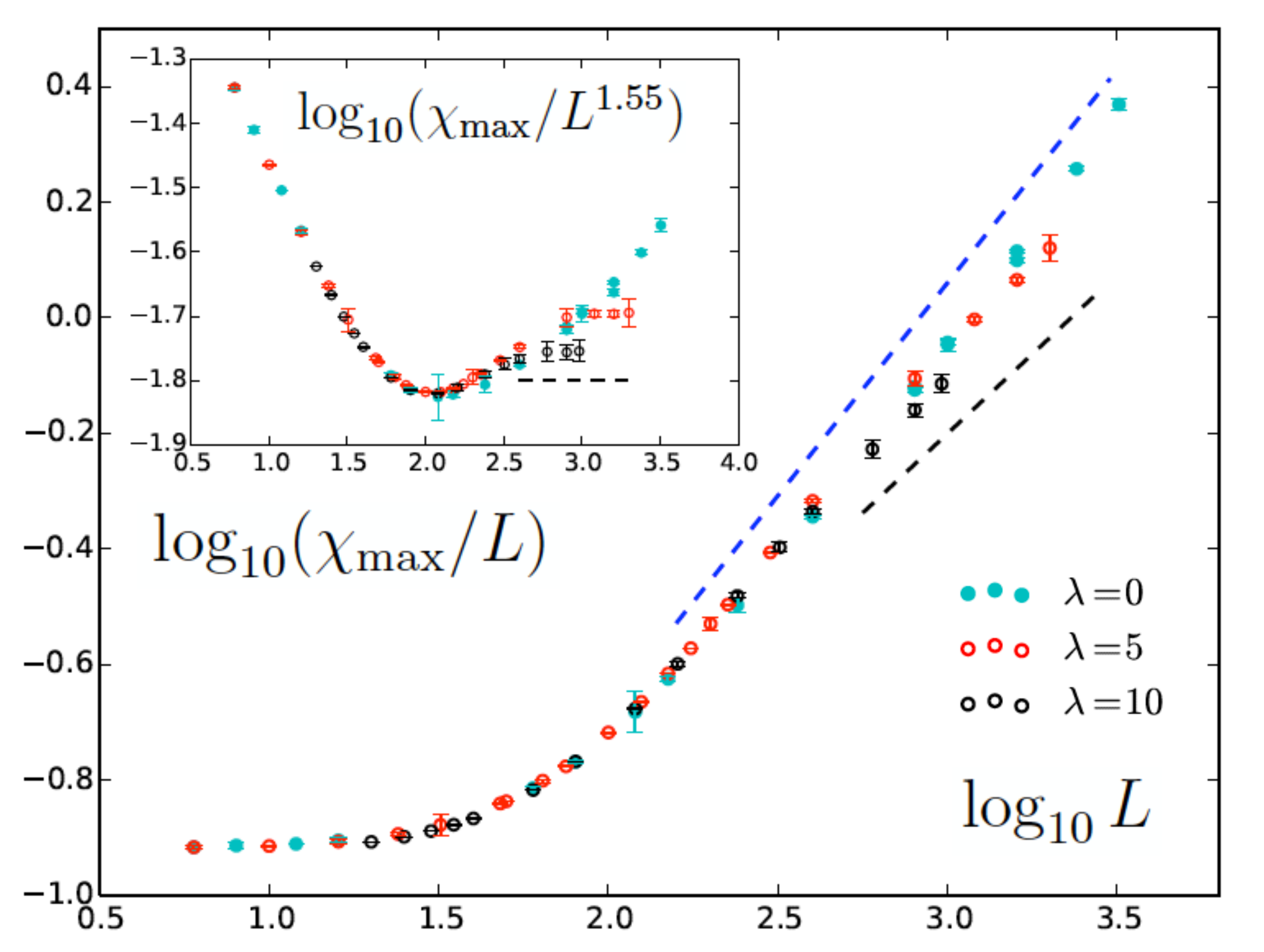}
\caption {Divergence of susceptibility peak value $\chi_{\max}$ with system size $L$, for different values of the advection coefficient $\lambda$. In the main panel $\chi_{\max}$ is rescaled by its small-$L$ power law $L^{1.0}$, whereas the inset shows $\chi_{\max}$ rescaled by $L^{\gamma/\nu}$ where $\gamma/\nu = 1.55$ is the estimate for the incompressible Toner-Tu (ITT) class. The black dashed line marks the $\chi_{\max} \sim L^{1.55}$ power law. The blue dashed line marks $\chi_{\max} \sim L^{1.737}$, which corresponds to the measured $\gamma/\nu = 1.737(1)$ of the dipolar XY class \cite{MC2010}. The ``ITT-5" ($\lambda = 5$) and ``ITT-10" ($\lambda = 10$) simulations are compared against ``ferro" ($\lambda = 0$), and ITT shows three finite-size scaling regimes where at smaller sizes, it behaves like ``ferro" and shows a first crossover from $\gamma/\nu \approx 1.0$ to $\gamma/\nu \approx 1.737$, and then at larger sizes, it deviates from ``ferro" at the second crossover to a smaller $\gamma/\nu$ with an upper bound of about $1.55$.}  
\label{fig:susmax}
\end{figure}

If size $L$ is large enough for the system to be in the scaling regime of a stable RG fixed point, conventional FSS analysis can be used to extract information about the associated critical exponents \cite{privman1990book,SRN1998,GC2010}. The basic underlying assumption is that finite-size effects near criticality can be accounted for by the ratio $L/\xi$, where $\xi$ is the correlation length of the infinite-size system. Then one can derive, for example, that the divergence with $L$ for the susceptibility peak value is associated with the exponent $\gamma/\nu$:
\begin{eqnarray}
\chi_{\max}(L) \propto L^{\gamma/\nu}.
\end{eqnarray} 
We fit the peak regions of the susceptibility transition curves using polynomials of degree 3 for $\ln\chi(a)$ to obtain the peak value $\chi_{\max}(L; \Delta x)$ and the peak location $a_{\rm c}^{\rm sus.}(L; \Delta x)$. Figure \ref{fig:susmax} shows the divergence of $\chi_{\max}$ with $L$, for the ``ITT-5", ``ITT-10" and ``ferro" simulations at various resolutions. 
The ITT data show three scaling regimes I, II and III as $L$ increases. In Regimes I and II, the ITT $\chi_{\max}(L)$ value almost equals the zero-advection ``ferro" value, and both show a crossover from the Regime-I exponent $\gamma/\nu \approx 1$ to the dipolar XY exponent $\gamma/\nu \approx 1.737$ \cite{MC2010}. In Regime III, advection comes into play and ITT deviates from ``ferro" to show a smaller $\gamma/\nu$.  This double-crossover behavior is consistent with the one-loop RG flows on the critical surface in $(4- \epsilon)$ dimensions \cite{CTL2015}: As size increases, following a certain flow line, the system first leaves the Gaussian fixed point, approaches the unstable fixed point of an isotropic ferromagnetic with dipolar interactions, and then deviates from it to go to the stable fixed point of ITT. %(\fref{fig:intro}). 
Our data confirm that the dipolar XY class is indeed a hyperbolic fixed point in 2D. 
Following this RG flow picture, we conclude that Regime III probably corresponds to the 2D ITT class. The data yield an estimate for the ITT fixed point: $\gamma/\nu = 1.55(8)$, or smaller if the current sizes are still too small to reveal the asymptotic scaling behavior. A smaller susceptibility peak value than the ``ferro" case indicates that advection tends to suppress the fluctuations of the global speed.

Here the role of the spatial resolution $\Delta x$ is worthy of note. In the RG picture, $\Delta x$ corresponds to the microscopic lengthscale at which the bare model parameters live, and thus one may expect that $\chi_{\max}(L; \Delta x)$ at different $\Delta x$ fall onto different curves with different crossover locations $L_1^{*}$ and $L_2^{*}$. This is true for Regime III: ``ITT-5m" and ``ITT-10" cross-over from Regime II to Regime III at different sizes, namely $L_2^{*} \approx 1000 \approx 10^{3.0} $ for ``ITT-5m" and $L_2^{*} \approx  501 \approx 10^{2.7} $ for ``ITT-10". Remarkably, our data show that, when all the bare model parameters except the advection coefficient remain the same, varying $\Delta x$ does not change the susceptibility peak value in Regimes I and II (see \ref{app:diffdx}); in figure \ref{fig:susmax} we have utilized this numerical observation and combined the data of different $\Delta x$ in Regimes I and II to show a more complete three-regime picture. The theory presented in \sref{sec:mft} can explain this $\Delta x$-independence for Regime I, but we still do not understand why varying $\Delta x$ does not change the location of the first crossover $L_1^{*} \approx 32 \approx 10^{1.5} $.

\begin{figure}
\centering
\includegraphics[width=8.0 cm]{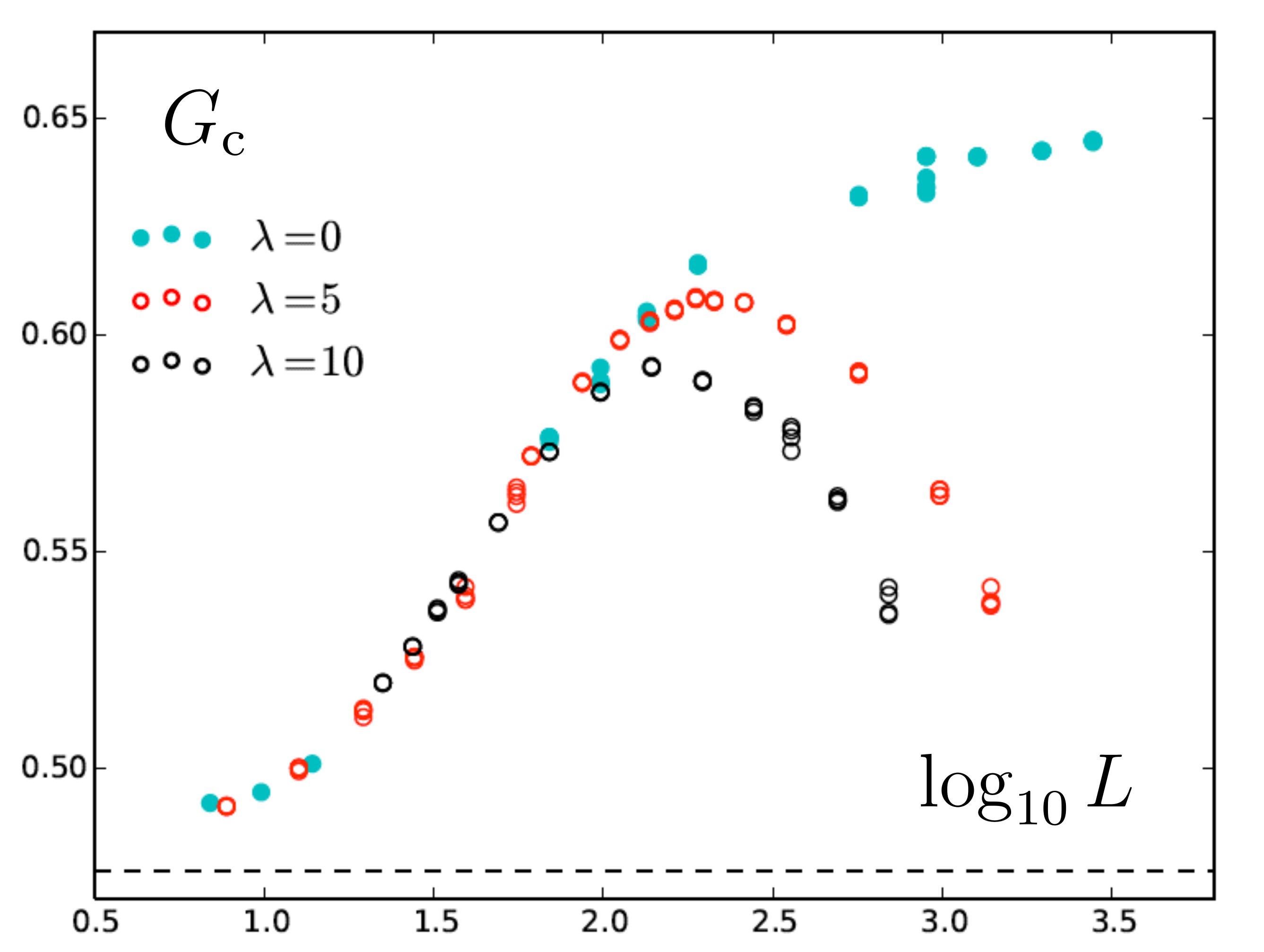}
\caption {Binder cumulant crossing value $G_{\rm c}$ versus  $\log_{10} L$, for different values of the advection coefficient $\lambda$. The dashed line marks the value $1-\pi/6$ as predicted by the weak-coupling critical theory (see (\ref{eq:GcWCCT})). Compared with ``ferro" ($\lambda = 0$), ``ITT-5" ($\lambda = 5$) and ``ITT-10" ($\lambda = 10$) show three regimes where at smaller sizes, ITT behaves like ``ferro" and makes a transition from a small plateau value close to $1-\pi/6$ to a large plateau value of about $0.64$ (close to the Binder cumulant value of the dipolar XY class as measured in \cite{MC2010}), and then at larger sizes, ITT deviates from ``ferro" to approach a smaller value with an upper bound of about $0.55$. At each size $L$, the same data are fitted using four different methods to locate the crossing points, and the vertical scattering of the four points provides a rough estimate of error.}  
\label{fig:GcL}
\end{figure}

In the conventional large-$L$ case without corrections to scaling, the Binder cumulant transition curves for different sizes (such as figure \ref{fig:trans-five}(c)) should cross at the same point $(-a_{\rm c}^{\infty}, G_{\rm c}^{\infty})$, where $a_{\rm c}^{\infty}$ is the infinite-$L$ critical point and $G_{\rm c}^{\infty}$ is a universal value associated to the universality class. Here in the presence of strong finite-size effect, the crossing points $(-a_{\rm c}^{\rm Binder}(L_1, ~L_2), ~G_{\rm c}(L_1,~L_2))$ between two consecutive sizes $L_1$ and $L_2$ change as $L_{12} \equiv \sqrt{L_1 L_2}$ increases. The crossing points are extracted by fitting each curve locally with a polynomial using a few points near crossing. To produce good-quality fits, we fit $G(a)$ for the ``ITT-5s" and ``ferro-s" simulations, and fit $\ln(2 -3 G)$ as a function of $a$ for the others. Four different fitting methods are used: (1) fitting a degree-2 polynomial using about four data points near crossing, (2) degree 2 using about five points, (3) degree 3 using about five points, (4) degree 3 using about seven points. The difference among fittings provides a rough estimate of error. In figure \ref{fig:GcL}, the measured Binder cumulant crossing values $G_{\rm c}$ are plotted against $\log_{10}(L_{12}) $. The value $G_{\rm c}$ is also found to be insensitive to $\Delta x$ in Regimes I and II (see \ref{app:diffdx}). We find three regimes, in agreement with the susceptibility peak divergence, though the two crossovers start at smaller sizes. In Regimes I and II, the ITT and ``ferro" simulations show the same behavior, and both increases from a small Regime-I value of less than 0.5, to approach a large value of about 0.64 that is consistent with the Monte Carlo measurements of the dipolar XY class \cite{MC2010}. The ITT data further display Regime III as advection comes in, where it deviates from ``ferro" and quickly decreases. This change in tendency appears at $L_2^{*} \approx 10^{2.1} \approx 126$ for ``ITT-10" and $L_2^{*} \approx 10^{2.3} \approx 200$ for ``ITT-5m". The data set an upper bound for the universal Binder cumulant $G_{\rm c}^{\infty}$ of the ITT class, namely $G_{\rm c}^{\infty} < 0.55$, a value much smaller than that of the dipolar XY class.

\begin{figure}
\centering
\includegraphics[width=12.0 cm]{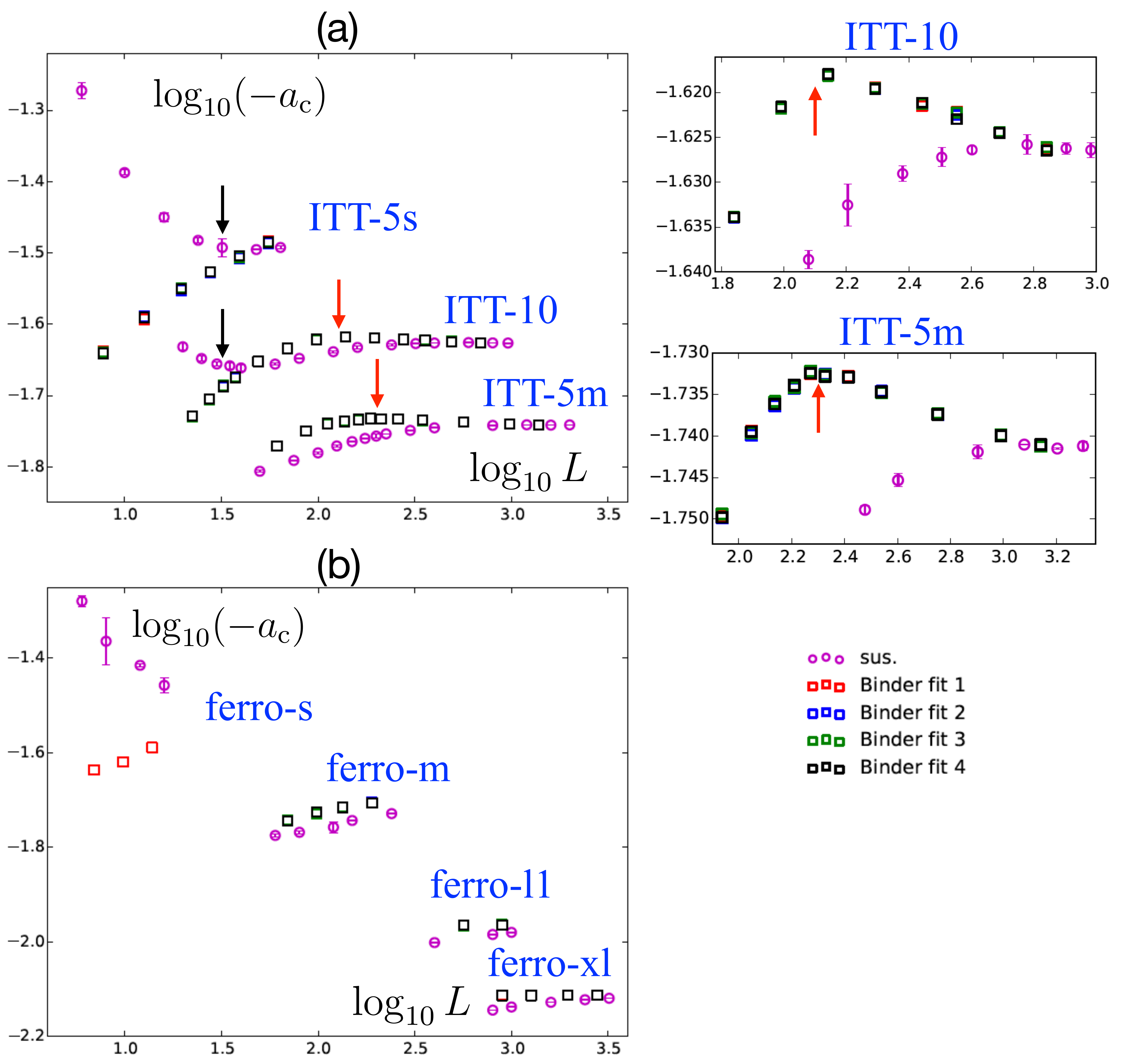}
\caption {Susceptibility peak location $-a_{\rm c}^{\rm sus.}(L)$ and Binder cumulant crossing location $-a_{\rm c}^{\rm Binder}(L_1, ~L_2)$ as functions of system size $L $ in $\log_{10}$-$\log_{10}$ scales, for (a) ITT simulations with the advection coefficient $\lambda = 5,~10$ and (b) ``ferro" simulations with $\lambda = 0$. The data points of different simulations group into different sets as labelled, as the location of the critical point is nonuniversal: For different values of model parameters, $(-a_{\rm c}^{\rm sus.})$ and $(-a_{\rm c}^{\rm Binder})$ approach different asymptotic values at infinite size. 
The double-crossover behavior as depicted in figures \ref{fig:susmax} and \ref{fig:GcL} is accompanied with tendency changes in the corresponding locations $(-a_{\rm c}^{\rm sus.})$ and $(-a_{\rm c}^{\rm Binder})$. In (a), black arrows are placed at the first crossover, and red arrows at the second crossover where the Binder cumulant crossing value $G_{\rm c}$ changes tendency. 
As shown in the magnified view to the right of (a), the critical points turn to the disordered side at the second crossover, which indicates that advection tends to stabilize order; such a tendency change is absent in (b). }  
\label{fig:mucL}
\end{figure}

With conventional leading-order corrections to scaling for large $L$, the susceptibility peak location $-a_{\rm c}^{\rm sus.}(L)$ and the Binder cumulant crossing location $-a_{\rm c}^{\rm Binder}(L_1,L_2 )$ should approach the same non-universal critical point $-a_{\rm c}^{\infty}$, with the same universal exponent $1/\nu$ but with typically different prefactors:
\begin{eqnarray}
|a_{\rm c}(L) - a_{\rm c}^{\infty} | \propto L^{-1/\nu},
\label{eq:delta-a}
\end{eqnarray}
where $a_{\rm c}(L)$ is $a_{\rm c}^{ \rm sus.}(L)$ or $a_{\rm c}^{\rm Binder}(L = \sqrt{L_1L_2})$. As shown in figure \ref{fig:mucL}, the directions in which $a_{\rm c}^{ \rm sus.}(L)$ and $a_{\rm c}^{\rm Binder}(L)$ approach the asymptotic value are different in the three regimes. In Regime I, $-a_{\rm c}^{\rm sus.}(L)$ decreases while $-a_{\rm c}^{\rm Binder}(L)$ increases with $L$. In Regime II, both increase. In Regime III, $-a_{\rm c}^{\rm Binder}(L)$ decreases with $L$ (note that the tendency change of the Binder cumulant crossing location coincides with that of the value $G_{\rm c}(L)$), and we find weak numerical evidence of $-a_{\rm c}^{\rm sus.}(L)$ also decreasing with $L$. The data show that when advection comes into play in Regime III, the finite-size critical points (and thus the limiting value $-a_{\rm c}^{\infty}$) shift to the disordered side, which supports that advection tends to stabilize order.

The three regimes are also observed in the steady-state statistics of the inhomogeneous modes near criticality. We further perform numerical simulations of ``ITT-5" and ``ferro" at the susceptibility peak location $a = a_{\rm c}^{\rm sus.}(L)$, and measure the steady-state averaged power spectrum $\langle { E({\bf k}, t) } \rangle$ defined as
\begin{eqnarray}
\langle{ E({\bf k}, t) } \rangle &\equiv& \frac{1}{N^4} \langle{{\bf v} ({\bf k}, t) \cdot {\bf v}^{*} ({\bf k}, t) }  \rangle,
\end{eqnarray}
where ${\bf v} ({\bf k}, t)$ is the discrete Fourier transform of the velocity field ${\bf v} ({\bf r}, t)$ (normalization convention defined in \ref{app:linear}). The factor $1/N^4$ is introduced so that for a given size $L$, the power spectra for different spatial resolutions $\Delta x$ fall onto the same curve. The power spectrum is expected to show power-law scaling in the $k \rightarrow  0$ limit 
\begin{eqnarray}
\langle  E({\bf k}, t)  \rangle  \sim k^{-(2-\eta)},
\end{eqnarray}
and based on the RG scaling hypothesis, the universal exponent $\eta = 2 -\gamma/\nu$. 

\begin{figure}
\centering
\includegraphics[width=10.0 cm]{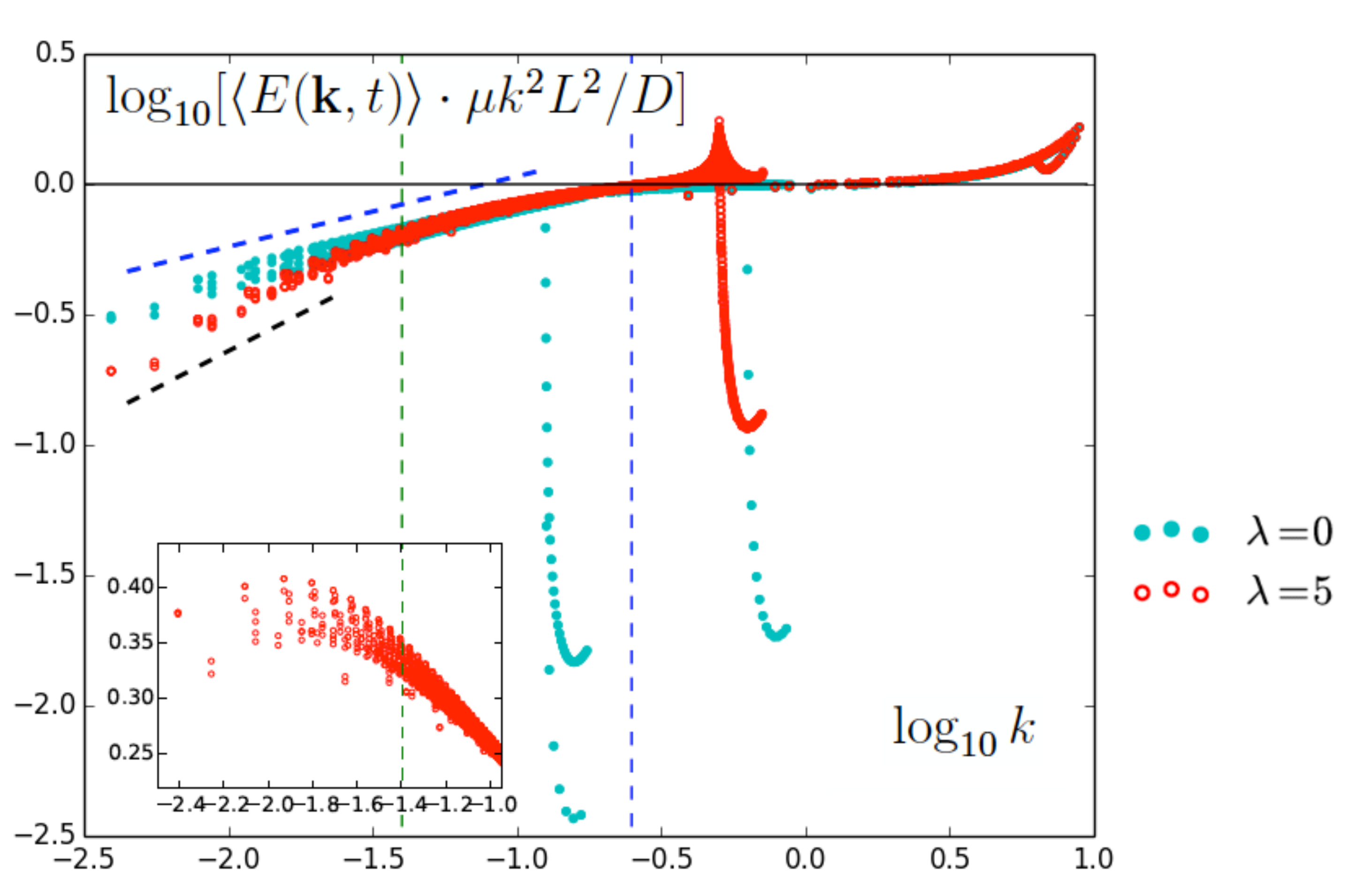}
\caption {Scaled average power spectrum $\langle { E({\bf k}, t) } \rangle \cdot \mu k^2 L^2/D$ as a function of the wavenumber $k \equiv |{\bf k}|$ in $\log_{10}$-$\log_{10}$ scales, for different values of the advection coefficient $\lambda$. Measurements are done at the susceptibility peak locations $a = a_{\rm c}^{\rm sus.}$. Compared to ``ferro" ($\lambda = 0$), the ``ITT-5" ($\lambda = 5$) power spectrum shows two $L$-independent crossovers at $ k_{\rm c1} \approx 0.25$ (vertical blue dashed line) and $ k_{\rm c2} \approx 0.04$ (green dashed line) respectively. At small scales where $k > k_{\rm c1} $, both ITT and ``ferro" follow the Gaussian solution (\ref{eq:power}) as denoted by the horizontal line $\log_{10}[\langle { E({\bf k}, t) } \rangle \cdot \mu k^2 L^2/D] = 0$; for $  k_{\rm c2} < k <  k_{\rm c1}$, both show the scaling $\langle { E({\bf k}, t) } \rangle \sim k^{-(2 - \eta)}$ with $\eta \approx 0.27$ (blue dashed line); at large scales with $k < k_{\rm c2}$, ITT deviates from ``ferro" and shows a larger exponent $\eta$ with a lower bound $\eta \approx 0.57$ (black dashed line). This is a $k$-space correspondence of the finite-size crossovers as shown in figures \ref{fig:susmax}-\ref{fig:mucL}. Plotted are ``ITT-5s" $L = 6$ and $L = 16$, ``ITT-5m" $L = 150$, $L = 400$ and $L = 1600$, ``ferro-s" $L = 6$ and $L = 16$, ``ferro-m" $L = 150$, ``ferro-l1" $L = 400$, and ``ferro-l2" $L = 1600$. Inset: $\log_{10} [\langle E({\bf k}, t)\rangle L^2 \cdot k^{2-\eta}]$ with estimated $\eta = 0.57$ is plotted against $\log_{10}k$ for ``ITT-5".}
\label{fig:powerEk}
\end{figure}

In figure \ref{fig:powerEk}, the solution of the Gaussian fixed point (\ref{eq:power}) is used as a basis, and the quantity $\langle { E({\bf k}, t) } \rangle \cdot \mu k^2 L^2/D$ is plotted against the wavenumber $k \equiv |{\bf k}|$ in $\log_{10}$-$\log_{10}$ scales for each ${\bf k}$-mode,  for sizes ranging from $L = 6$ (Regime I) to $L = 1600$ (Regime III). The Gaussian solution is marked by the horizontal black line. The error can be estimated from the vertical scattering of points. We observe that the scaled power spectrum data $\langle E({\bf k}, t) \rangle L^2$ fall onto an $L$-independent curve (see \ref{app:pow-diffL}). The data points deviate from this curve at the largest wavenumbers in each simulation, but this abnormal tail is only a numerical artifact due to the pseudo-spectral scheme at finite resolution $\Delta x$ (also see \ref{app:pow-diffL}). 
We observe that the ``ITT-5" power spectrum shows two size-independent crossovers, one at $ k_{\rm c1} \approx 0.25$ and the other at $ k_{\rm c2} \approx 0.04$. For wavenumbers larger than $ k_{\rm c2}$, ITT behaves the same as ``ferro", and both show a crossover at $k_{\rm c1}$ from the larger-$k$ Gaussian behavior to the smaller-$k$ dipolar XY scaling. At small wavenumbers, the ``ferro" power spectrum follows a power law with $\eta = 0.27(8)$, which gives $\gamma/\nu = 2 - \eta =  1.73(8)$ in agreement with $\gamma/\nu = 1.737(1)$ for the dipolar XY class\cite{MC2010}. Advection only affects wavenumbers smaller than $ k_{\rm c2}$, and in this regime, ITT deviates from ``ferro" to show a larger $\eta$. The exponent $\eta$ for the ITT class is measured to be $\eta = 0.57 (11)$, or larger if the smallest-$k$ data are not yet in the asymptotic scaling regime. Note that this three-regime picture of the eddy power spectrum as wavenumber $k$ decreases matches with the aforementioned three regimes of the mean-flow statistics as size $L$ increases. The smallest nonzero wavenumber $k_{\rm min} = 2 \pi/L $ decreases with $L$. When $k_{\rm min}$ decreases to $k_{\rm c1}$, the corresponding size $L \approx 25$ is close to $L_1^{*} \approx 32$ of the first crossover in the mean-flow statistics; when it reaches $k_{\rm c2}$, the size $L \approx 157$, close to $L_2^{*} \approx 200$ when the Binder cumulant crossing value and location change tendency at the second crossover.         
 
\begin{figure}
\centering
\includegraphics[width=9.0 cm]{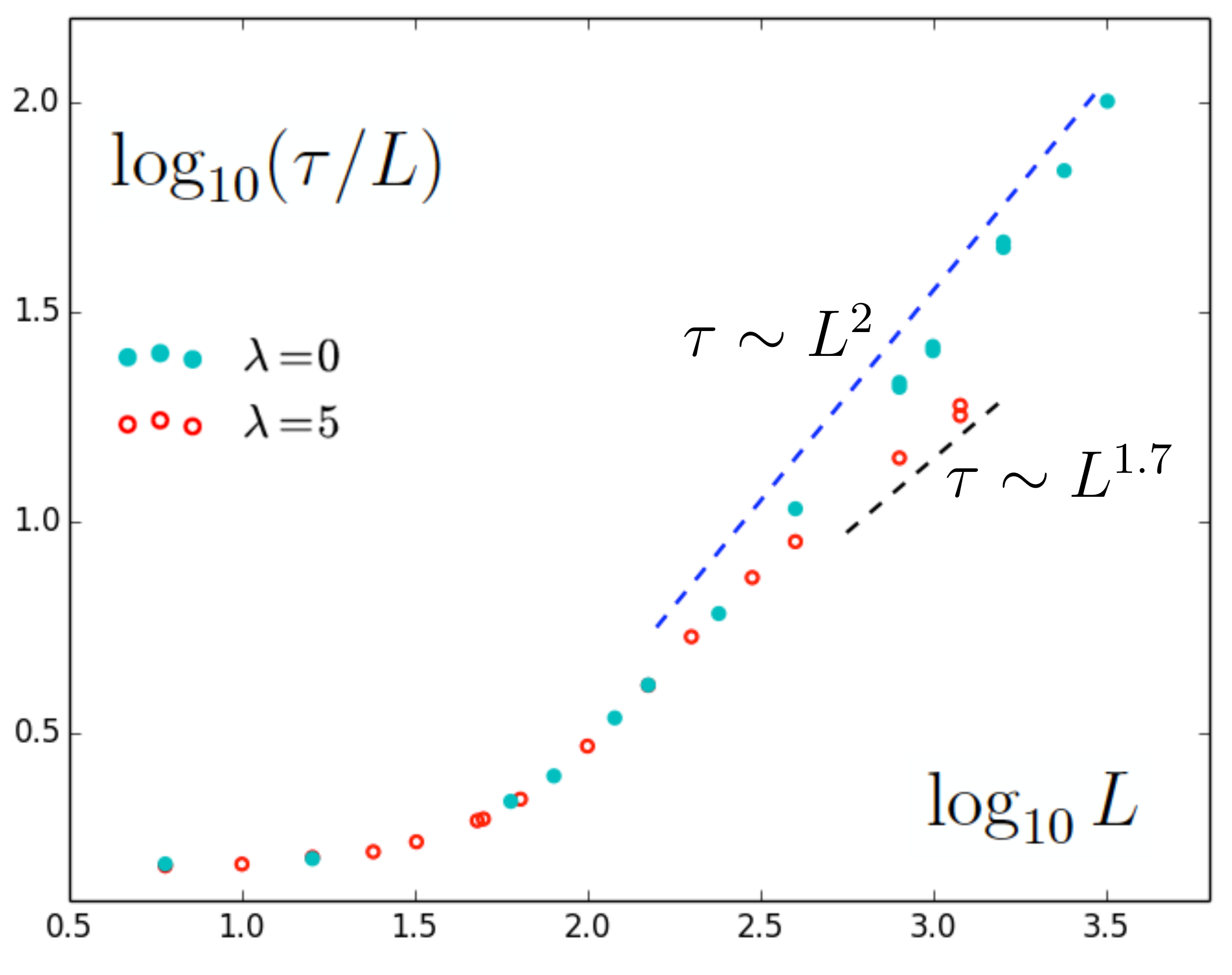}
\caption {Divergence of autocorrelation time $\tau$ with system size $L$, for different values of the advection coefficient $\lambda$. Measurements are done at the susceptibility peak locations $a = a_{\rm c}^{\rm sus.}$. Here $\tau$ is rescaled by its small-$L$ power law $L^{1.0}$. Compared to ``ferro" ($\lambda = 0$), ``ITT-5" ($\lambda = 5$) exhibits three scaling regimes: At smaller sizes, ITT behaves like ``ferro" and they show a crossover from the scaling $\tau \sim L^z$ with the dynamic exponent $z \approx 1.0$ to the diffusive scaling with $z \approx 2.0$; at large sizes as advection comes into play, the transport of fluctuations in ITT becomes faster than the diffusive dynamics in the equilibrium ``ferro" system, and we find a smaller $z$ with an upper bound of about $1.7$. }  
\label{fig:tau}
\end{figure}

After presenting the static scaling results for both the mean flow and the eddies, we now turn to the dynamic scaling of the mean flow. We study critical slowing down by measuring the global-speed autocorrelation function 
\begin{eqnarray}
C(t) \equiv \frac{ \langle c(T) c(t + T)\rangle_T - [\langle c(T) \rangle_T]^2}{ \langle c^2(T) \rangle_T - [\langle c(T) \rangle_T]^2}
\end{eqnarray}
at the susceptibility peak location $a = a_{\rm c}^{\rm sus.}(L)$. The autocorrelation time $\tau$ is obtained by fitting $C(t) \sim e^{-t/\tau}$ for large lag time $t$. In the conventional FSS analysis, the autocorrelation time grows with size as
\begin{eqnarray}
\tau(L) \propto L^z,
\end{eqnarray}
where $z$ is the dynamic critical exponent. Figure \ref{fig:tau} shows the divergence of $\tau$ for the ``ITT-5" and ``ferro" simulations at various resolutions, and we find the three-regime scaling behavior similar to that of the susceptibility peak value $\chi_{\max}(L)$. The autocorrelation time $\tau$ is also observed to be insensitive to spatial resolution $\Delta x$ in Regimes I and II. Details can be found in \ref{app:tau-related}. The dynamic exponent $z \approx 1$ in Regime I, whereas the transport of fluctuations across the system is diffusive in Regime II with $z \approx 2$ and it becomes faster than diffusive in Regime III with $z < 2$. The data yield a crude estimate, $z \approx 1.7 $ or smaller, for the ITT class. Compared to the $\chi_{\max}(L)$ behavior, the first crossover size is approximately the same, namely $L_1^{*} \approx 32 \approx 10^{1.5}$, but the second crossover happens at a smaller size, $L_2^{*} \approx 398 \approx 10^{2.6}$ for ``ITT-5m"; the dynamic scaling is more sensitive to the effect of advection. 

\subsection{Summary and discussion: regimes, renormalization group and activity \label{subsec:summary}}

The critical behavior of the ITT model show three regimes as size increases. This is revealed in various aspects: first, in the static scaling of the mean flow in terms of the susceptibility peak divergence, the Binder cumulant convergence and the trends of the finite-size critical points; second, in the static scaling of the eddies in terms of the power spectrum; and last, in the dynamic scaling of the mean flow in terms of the autocorrelation time divergence. ITT behaves the same as the zero-advection ``ferro" system in Regimes I and II. Regime II is identified with the equilibrium universality class of the dipolar XY model. In Regime III advection further comes into play, and this regime corresponds to the 2D ITT class. The numerical estimates and bounds of the Binder cumulant and some exponents of the three regimes are summarized in table \ref{table:regimes}. Since accurate determination of the infinite-size critical point $a_{\rm c}^\infty$ is currently beyond our reach, the estimates of $1/\nu$ or $\beta/\nu$ cannot be obtained here.

%\begingroup
%\setlength{\tabcolsep}{8pt} % Default value: 6pt
\begin{table}
\caption{\label{table:regimes}Numerical estimates of the exponent ratio $\gamma/\nu$, the exponents $\eta$ and $z$ and the Binder cumulant $G_{\rm c}^\infty$ in the three regimes, compared with the one-loop renormalization group (RG) predictions for the 2D ITT class \cite{CTL2015}. Whether these universal quantities are measured from the $k = 0$ mode or the $k > 0$ modes is explicitly indicated.}
\begin{indented}
\item[]\begin{tabular}{ @{}c l l l l l  }
\br
 Regimes &  I &  II (Dipolar XY)  &  III (2D ITT)  & RG Range & RG Bound  \\
 \mr
$\gamma/\nu$ ($k = 0$) &  $\approx 1.0$ & $\approx 1.737$ &   1.55(8) or smaller &  1.21(25) &  $<1.73$ \\
$\eta$ ($k > 0$)& $\approx 0$ & 0.27(8) &   0.57(11) or larger &  0.79(25) & $> 0.27$ \\
 $z$ ($k = 0$) & $\approx 1.0$ & $\approx 2.0$  &  $\approx 1.7$ or smaller   &  1.21(25)  & $<1.73$ \\
 $G_{\rm c}^\infty$ ($k = 0$) & $< 0.5$ & $\approx 0.64$ &  $< 0.55$ & & \\
 \br
 \end{tabular}
 \end{indented}
 \end{table}
% \endgroup

The numerical estimates of the 2D ITT exponents are compared with the one-loop RG predictions also in table \ref{table:regimes}. Since the RG analysis is to linear order in $\epsilon = 4 -d$, it is subtle how to obtain numerical values of the exponents in $d = 2$. Here we use two methods. The method used in \cite{CTL2015} for $d = 3$ utilizes both the $\epsilon$-expansion and the exact scaling relations to yield a range of values for each exponent, and here it is directly applied to $d = 2$ and shown as ``RG range". This quantitative estimate in 2D may not be reliable, so in the second method, we only claim that the exponents should be further away from their mean-field values than they are in $d = 3$, and the result is shown as ``RG bound". Note that at the one-loop level, the relation $z = \gamma/\nu$ holds, because the absence of diagrammatic correction to noise strength leads to the scaling relation $z = 2 \chi + d $ (with $\chi $ being the ``roughness" exponent in the RG analysis, not to be mixed with the susceptibility); this relation becomes invalid at two-loop or higher orders. Valid to all loops are the relations $\eta = 2 - \gamma/\nu$, and $\eta =  2 -d - 2 \chi$ based on the RG scaling hypothesis. We find that the crude one-loop RG estimates for 2D ITT are consistent with our numerics. We also note that the Kardar-Parisi-Zhang (KPZ) exponents that appear in the 2D ITT flocking phase \cite{CLT2016} are not related to these critical exponents, because the velocity patterns near criticality abound with vortices in Regime III (see \fref{fig:intro}) and the mapping to KPZ thus breaks down.

The activity-induced crossover from equilibrium to off-equilibrium deserves more discussion. Based on simple power counting, ignoring the diagrammatic corrections, we find that the advective nonlinearity grows exponentially in Regime II and will become relevant as size further increases (see \ref{app:scaling_argument}). This agrees with the observation of Regime III following Regime II. The effects of advection (or, activity) are revealed in various aspects. Qualitatively, advection makes the finite-size critical points change tendency to shift to the disordered side, which shows that advection tends to stabilize collective motion. Quantitatively, the dynamic exponent $z$ is changed from the diffusive value to a smaller value, which suggests that advection causes more efficient mechanisms of information transfer in velocity fluctuations; advection also leads to a smaller exponent ratio $\gamma/\nu$ and a larger exponent $\eta$ as related by a scaling relation, showing that advection suppresses large-scale fluctuations. These observations qualitatively agree with the one-loop RG predictions in $(4-\epsilon)$ dimensions. RG predicts $z = 2 + {\cal O}(\epsilon^2)$ for the dipolar ferromagnet, and a smaller $z = 2 - 31\epsilon/113 +  {\cal O}(\epsilon^2)$ for ITT \cite{CTL2015}; both fixed points satisfy $z = \gamma/\nu + {\cal O}(\epsilon^2)$ as noise strength receives no diagrammatic correction, and this relation between the dynamic and static scaling suggests that for both equilibrium and non-equilibrium, faster internal processes lead to smaller fluctuations at large scales, hence more stable long-range order. Future studies may clarify the non-equilibrium mechanism by which activity expedites information transfer. Moreover, the universal Binder cumulant decreases drastically as advection comes in, which reflects a significant change in the probability distribution function of the global speed; this also remains to be understood.

Finally, we note that the scaling exponents in Regime I, an equilibrium regime shared by the dipolar XY model and the ITT model, do not agree with the RG predictions around the Gaussian fixed point.
%. The flow trajectory implies that Regime I corresponds to the Gaussian fixed point, in agreement with the observed power spectrum, but the Gaussian exponents $z = \gamma/\nu = 2$ disagrees with the observed $z \approx \gamma/\nu \approx 1.0$. As we show below, the key to understanding this seemingly anomalous behavior lies in the near-decoupling between the $k = 0$ mode and the $k > 0$ modes when the nonlinear effects are severely constrained by system size; the $k > 0 $ modes are Gaussian, whereas the $k = 0 $ scaling exponents are mean-field. By integrating out the large-$k$ shells, RG addresses the $k \rightarrow 0^{+}$ limit but does not deal with the $k = 0$ dynamics, and thus RG only provides a partial description of Regime I. 
The discrepancy is examined in the section below where a reduced hydrodynamic theory, the WCCT, is shown to predict correctly the observed exponents as well as other aspects of the transition in Regime I.

\section{Weak-coupling critical theory for Regime I \label{sec:mft}}

We start by assuming that the eddy modes are described by the Gaussian fixed point, in agreement with the RG flow. Thus the exact dynamics (\ref{eq2}) of eddies is now simplified into the critical linear dynamics
\begin{eqnarray}
 \frac{\partial \omega}{\partial t} &=&  \mu \nabla^2 \omega + \eta({\bf r}, t).\label{WCCT-eq2}
\end{eqnarray}
The corresponding statistics is solved in \ref{app:linear}. Taking into account the discrete and finite-size nature of numerical models, we find an averaged power spectrum
\begin{eqnarray}
\overline{ E({\bf k}, t) } &\equiv& \frac{1}{N^4} \overline{{\bf v} ({\bf k}, t) \cdot {\bf v}^{*} ({\bf k}, t) } =     \frac{  D  }{\mu k^2 L^2 } ,  \label{eq:power}
\end{eqnarray}
for any discrete wavenumber $k>0$. Here the steady-state average in the reduced hydrodynamics is denoted by $\overline{\cdots}$, to be distinguished from that of the full dynamics $\langle \cdots \rangle$. 
Equation (\ref{eq:power}) has already been shown in figure \ref{fig:powerEk} to agree with numerical simulations in Regime I. 
Another consequence of this assumption is that advection is completely neglected, because as mentioned before, the advective nonlinear terms in the full model only appear in the eddy dynamics due to Galilean invariance. Thus the theory predicts that Regime-I statistics are independent of the values of the advection coefficient $\lambda$, which agrees with numerical simulations.

\begin{figure}
\centering
\includegraphics[height=5.7 cm]{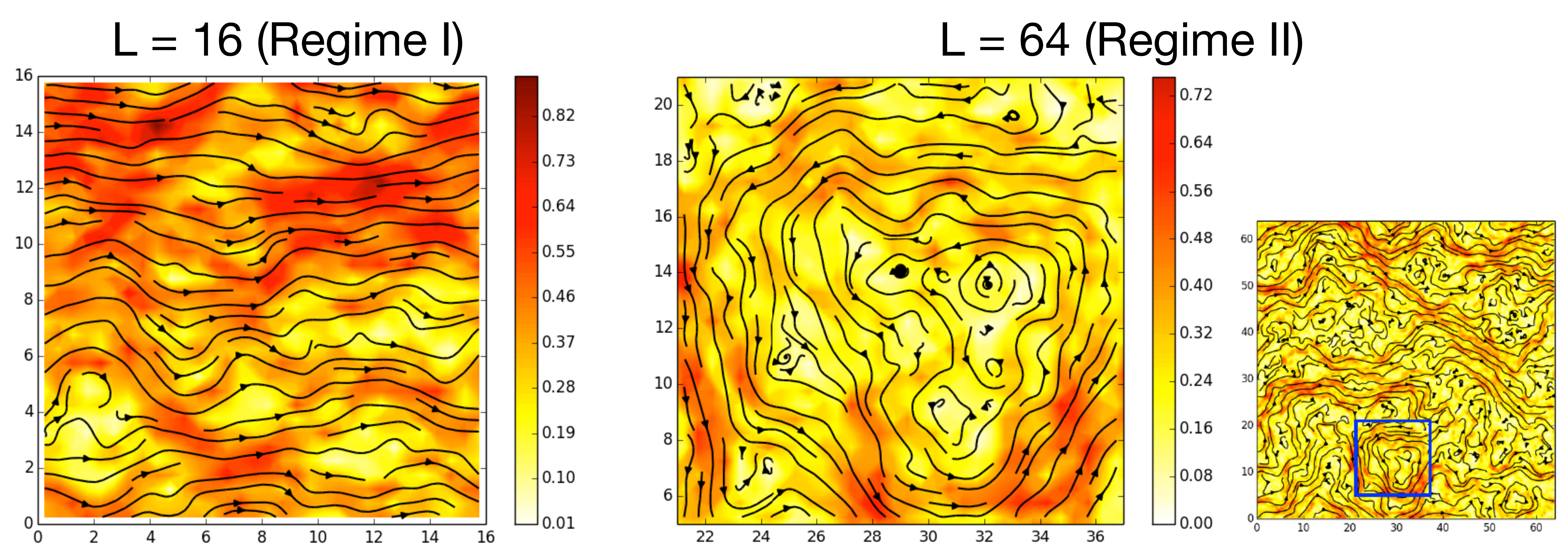}
\caption {Typical steady-state snapshots of velocity field at the susceptibility peak locations $a = a_{\rm c}^{\rm sus.}$, for the ``ITT-5s" simulations in Regime I and Regime II. The streamlines are plotted, and the local speed $|{\bf{v}}|({\bf r}, t)$ is shown in color. The crossover from Regime I to Regime II is accompanied with a morphological change in typical velocity field patterns. The typical patterns in Regime I are dominated by disorder patches and jets, but no vortex, as exemplified by the $L = 16$ snapshot. As size increases and the system enters Regime II, vortices appear and patterns become more isotropic, as exemplified by the $L = 64$ case; the whole snapshot is given on the right, and the $16\times16$ vortex region inside the blue square is shown on the left for comparison with $L = 16$. This hints at the dominance of the mean flow in Regime I. The global velocity is in the $\hat{\bf x}$ direction for both snapshots.}
\label{fig:snapshots}
\end{figure}

Numerics further reveal a morphological change in typical velocity field patterns near criticality when the system crosses-over from Regime I to Regime II, which hints at the dominance of the mean flow in Regime I. This is illustrated in \fref{fig:snapshots}. For systems in Regime I near the susceptibility peak location, the velocity field is dominated by patches of disorder domains where the local speed almost vanishes, and the jets of nonzero velocity that pass by them; the jets maintain a global direction and rarely wind around to form vortices. By contrast, as the system enters Regime II, vortices appear and patterns become more isotropic. This observation suggests that near criticality, the mean flow ${\bf c}(t)$ must be large enough compared to the eddies ${\bf v}'({\bf r}, ~t)$ to keep the system in Regime I. Such a dominance could imply weak coupling of the dynamics, as the kinetic energy in the mean flow may be easily transmitted into the eddies if they are strongly coupled. We thus assume that the eddies contribute to the mean-flow dynamics only through statistical averages. The exact dynamics (\ref{eq1}) is now approximated as
\begin{eqnarray}
 \fl \frac{d {\bf c}}{dt} &=& (-a - b |{\bf c}|^2 - b ~\overline{ \langle |\nabla \psi|^2 \rangle_{\bf r}}~) {\bf c} - b~ \overline{ \langle |\nabla \psi|^2 {\bf v}' \rangle_{\bf r} } -  2 b~  \overline{ \langle {\bf v}' ({\bf c} \cdot {\bf v}' )\rangle_{\bf r} }  + \langle {\bf f} \rangle_{\bf r} , \label{WCCT-eq1}
 \end{eqnarray}
where the terms involving eddies have been replaced by their steady-state averages based on (\ref{WCCT-eq2}). We term this reduced hydrodynamic theory (\ref{WCCT-eq2}) and (\ref{WCCT-eq1}) the WCCT.

The dynamics of the mean flow can be written in a simpler form. Since $\overline{ \langle |\nabla \psi|^2 {\bf v}' \rangle_{\bf r} }  = 0$ and $\overline{ \langle {\bf v}' ({\bf c} \cdot {\bf v}' )\rangle_{\bf r} } = \frac{1}{2} \overline{ \langle |\nabla \psi|^2 \rangle_{\bf r}} ~{\bf c}$ (see \ref{app:linear} for derivation), equation (\ref{WCCT-eq1}) can be rewritten as
\begin{eqnarray}
\frac{d {\bf c}}{dt} &=& (-a_{\rm R} - b |{\bf c}|^2 ) {\bf c} +  {\boldsymbol\zeta}(t) =  - \frac{d U}{d {\bf c}}  +  {\boldsymbol\zeta}(t), \label{WCCT-eq1-2}
\end{eqnarray}
where the eddy fluctuations result in a constant shift $a_0$ in the linear coefficient,
\begin{eqnarray}
a_{\rm R} &\equiv& a - a_0, \\
a_0 & \equiv& - 2 b ~\overline{ \langle |\nabla \psi|^2 \rangle_{\bf r}}.
\end{eqnarray}
The new noise term
\begin{eqnarray}
{\boldsymbol\zeta}(t) \equiv \langle {\bf f}({\bf r}, t) \rangle_{\bf r}, 
\end{eqnarray}
and the potential 
\begin{eqnarray}
U({\bf c}) = \frac{a_{\rm R}}{2} |{\bf c}|^2 + \frac{b}{4} |{\bf c}|^4. 
\end{eqnarray} 
This is a 2D random walk in potential $U({\bf c})$. The statistical property of the noise ${\boldsymbol\zeta}(t)$ can be understood more clearly as follows. In a discretized form, the noise $  f_i({\bf r}, t)$ can be implemented as an independently-drawn Gaussian random number at each lattice point at each time step with zero mean and variance $2D/[(\Delta x)^2 \Delta t]$. For a large lattice, according to the central limit theorem, the spatially averaged noise $\langle {f_i}\rangle_{\bf r}$ is an independently-drawn Gaussian random number at each time step with zero mean and variance $2D/(L^2 \Delta t)$. In the $\Delta t \rightarrow 0$ limit,
\begin{eqnarray}
\overline{ \zeta_i(t) \zeta_{j}(t') } = 2 \left( \frac{D}{L^2} \right) \delta_{ij} \delta(t - t'),
\label{eq:noise-correlator}
\end{eqnarray}
where the Cartesian coordinates $i, ~j = x,~y$.
Note that in the thermodynamic limit $L \rightarrow \infty$, the noise vanishes and the deterministic dynamics has a critical point at $a_{\rm R} = 0$.

%%% WQ: general d-dim. WCCT part and z = d/2 are now removed.
%Before presenting an explicit solution of equation (\ref{WCCT-eq1-2}), we first consider a scaling analysis, and replace $L^2$ term in (\ref{eq:noise-correlator}) by $L^d$ in $d$ dimensions. Carrying out the scaling transformation ${\bf c}\rightarrow c_0{\bf c}$, $t\rightarrow t_0t$, and ${\boldsymbol{\zeta}}\rightarrow \zeta_0{\boldsymbol{ \zeta}}$, equations (\ref{WCCT-eq1-2}) and (\ref{eq:noise-correlator}) are brought to dimensionless form with the choice
%\begin{equation}
   % c_0=\Bigl(\frac{D}{b}\Bigr)^{1/4}L^{-d/4},\quad t_0=(bD)^{-1/2}L^{d/2},\quad\zeta_0=b^{1/4}D^{3/4}L^{-3d/4},
   % \label{eq:WCCT-scaling}
%\end{equation}
%while the term $a_{\rm R}$ is replaced by $a_{\rm R}t_0=a_{\rm R}(bD)^{-1/2}L^{d/2}$.
%Consequently, for a $d$-dimensional system in Regime I, the typical strength of the velocity field and the typical relaxation time both exhibit power-law dependence on the linear system size $L$ at the critical point $a_{\rm R}=0$, with respective exponents $\beta/\nu=d/4$ and $z=d/2$. 

%To obtain the full solution of equation (\ref{WCCT-eq1-2}), we 
We start with the Fokker-Planck equation for the probability distribution function $P({\bf c}, t)$,
\begin{eqnarray}
\frac{\partial P({\bf c}, t)}{ \partial t} &=& \sum\limits_{i = 1}^{2} \frac{\partial }{\partial c_i} \left[ \left(  \frac{\partial U}{\partial c_i}\right) P({\bf c}, t) \right] + \frac{D}{L^2} \sum\limits_{i = 1}^{2} \frac{\partial^2}{ \partial c_i^2}P({\bf c}, t) \nonumber\\
&=& \sum\limits_{i = 1}^{2} \frac{\partial }{\partial c_i} \left[( a_{\rm R} + b |{\bf c}|^2) c_i P({\bf c}, t) \right] + \frac{D}{L^2} \sum\limits_{i = 1}^{2} \frac{\partial^2}{ \partial c_i^2}P({\bf c}, t). \label{eq:FP}
\end{eqnarray}
In \ref{app:scaling}, we show that the statistics of the steady-state distribution satisfy the exact FSS forms
\begin{eqnarray}
\overline{c}(a, L) &=& L^{-1/2} \cdot F_1 [(a - a_0)  L] , \label{eq:cscalingformWCCT}\\
\chi(a, L) &=& L \cdot  F_2 [(a - a_0)  L], \label{eq:chiscalingformWCCT}\\
G(a, L) &=& F_3[(a - a_0) L], \label{eq:GscalingformWCCT}
\end{eqnarray}
where the scaling functions $F_1$, $F_2$ and $F_3$ are also given explicitly.
%The agreement between these scaling relations and our dimensional analysis (\ref{eq:WCCT-scaling}) is evident.

Despite that these scaling forms are given by a theory that addresses small-$L$ systems, they resemble the conventional FSS hypothesis for large-$L$ systems. The latter for the mean order parameter on the ordered side is
\begin{eqnarray}
\langle {c} \rangle (a, L) &=& L^{-\beta/\nu} \cdot \tilde{F}_1(|a - a_{\rm c}^\infty|L^{1/\nu}),
\end{eqnarray}
and that for the susceptibility is
\begin{eqnarray}
\chi(a, L )&=& L^{\gamma/\nu} \tilde{F}_2(|a - a_{\rm c}^\infty|L^{1/\nu} ),
\end{eqnarray}
where $\tilde{F}_1$ and $\tilde{F}_2$ are scaling functions. There is one major difference:
Here using (\ref{eq: vvcor}), the constant 
\begin{eqnarray}
\fl a_0 \equiv  - 2 b ~\overline{ \langle |\nabla \psi|^2 \rangle_{\bf r}}  &\approx&  -    \frac{2 b D}{ \mu L^2 }   {\sideset{}{'}\sum\limits_{n_{\rm x}, n_{\rm y} = -N/2}^{N/2 -1}}    \frac{1}{ k^2}  =    - \frac{bD}{2 \pi^2 \mu} {\sideset{}{'}\sum\limits_{n_{\rm x}, n_{\rm y} = -N/2}^{N/2 -1}}    \frac{1}{n_{\rm x}^2 + n_{\rm y}^2} , 
  \label{eq:aoWCCT}
\end{eqnarray}
not only depends on the parameters $\{b, \mu, D \}$ as the non-universal critical point $a_{\rm c}^\infty$ does, but also depends on $N$ (or, on system size $L$ for a given resolution $\Delta x$). The prime on the summation means to exclude $(n_{\rm x}, n_{\rm y}) = (0,0)$. The large-$N$ behavior of the sum ${\sideset{}{'}\sum\limits_{n_{\rm x}, n_{\rm y} = -N/2}^{N/2 -1}}    \frac{1}{n_{\rm x}^2 + n_{\rm y}^2}$ can be estimated by taking the continuum approximation and further approximating the square domain as a circle,
\begin{eqnarray}
\fl {\sideset{}{'}\sum\limits_{n_{\rm x}, n_{\rm y} = -N/2}^{N/2 -1}}    \frac{1}{n_{\rm x}^2 + n_{\rm y}^2} \sim \int_{1}^{N/2} 2 \pi r \rmd r~ \frac{1}{r^2} + {\rm higher~orders} = 2 \pi \ln (N/2) + {\rm higher~orders}. 
\end{eqnarray}
Despite that the $N$-dependent critical point $a_0(N)$ now replaces the constant $a_{\rm c}^\infty$, the exponents are identified as
\begin{eqnarray}
\gamma/\nu &=& 1,\\
1/\nu &=& 1, \\
\beta/\nu &=& 1/2.
\end{eqnarray}

For systems at $a = a_0(N)$, the steady-state statistics can be analytically calculated. The steady state is given by the Boltzmann distribution
\begin{eqnarray}
P_{\rm s.s.}({\bf c}) = \frac{1}{Z} \rme^{-L^2 U/D},
\end{eqnarray}
where the partition function 
\begin{eqnarray}
Z = \int \rmd^2{\bf c} ~\rme^{-L^2 U/D} = \int \rmd^2{\bf c} ~\exp \left(-\frac{b L^2}{4 D}  c^4 \right).
\end{eqnarray}
The mean order parameter, the susceptibility and the Binder cumulant are
\begin{eqnarray}
 \overline{ c} &=& \frac{\int_0^\infty \rmd c ~2 \pi c^2 \exp \left( -\frac{b L^2}{4D}  c^4 \right) }{\int_0^\infty \rmd c ~2 \pi c \exp \left(-\frac{b L^2}{4D}  c^4 \right)} \nonumber\\
 &=& \frac{\Gamma(3/4)}{\sqrt{\pi} } \left( \frac{4D}{b} \right)^{1/4} L^{-1/2} = L^{-1/2} \cdot F_1(0) , \label{eq:F10}\\
\chi &=&   \left\{  \frac{1}{\sqrt{\pi}}   -  \frac{[\Gamma(3/4)]^2}{\pi}   \right\}  \left( \frac{4D}{b } \right)^{1/2} L = L \cdot F_2(0), \label{eq:F20}\\
G_{\rm c} &=&  1 - \frac{\pi}{6 } \approx 0.4764 = F_3(0) \label{eq:GcWCCT}.  
\end{eqnarray}

The dynamic exponent $z$ can be inferred by considering a system at $a = a_0(N)$ in which initially the velocity ${\bf v}({\bf r}, 0) = {\bf 0}$, and the ensemble averaged global speed $\bar{ c}$ then grows with time and finally saturates. When the global velocity ${\bf c}$ is small at early times, the potential term in (\ref{WCCT-eq1-2}) is negligible compared to the noise, and ${\bf c}(t)$ behaves like the position vector in a 2D free random walk. The initial evolution is governed by the Fokker-Planck equation 
\begin{eqnarray}
\frac{\partial P({\bf c}, t)}{ \partial t} &=&\frac{D}{L^2} \sum\limits_{i = 1}^{2} \frac{\partial^2}{ \partial c_i^2}P({\bf c}, t),
\end{eqnarray}
with the solution 
\begin{eqnarray}
P({\bf c}, t) = \frac{1}{\pi A t  } \rme^{- c^2/(A t)}, 
\end{eqnarray}
where $A \equiv {4D}/{L^2}$. The mean 
\begin{eqnarray}
\overline{ c}(t) &=&  \frac{\sqrt{ \pi Dt}}{L},
\end{eqnarray}
which will saturate to the steady-state value (\ref{eq:F10}) at time  $t_{\rm sat.} \sim L^z$ with the dynamic exponent 
\begin{eqnarray}
z = 1. \label{eq:zWCCT}
\end{eqnarray}

\begin{figure}
\centering
\includegraphics[width=15.0 cm]{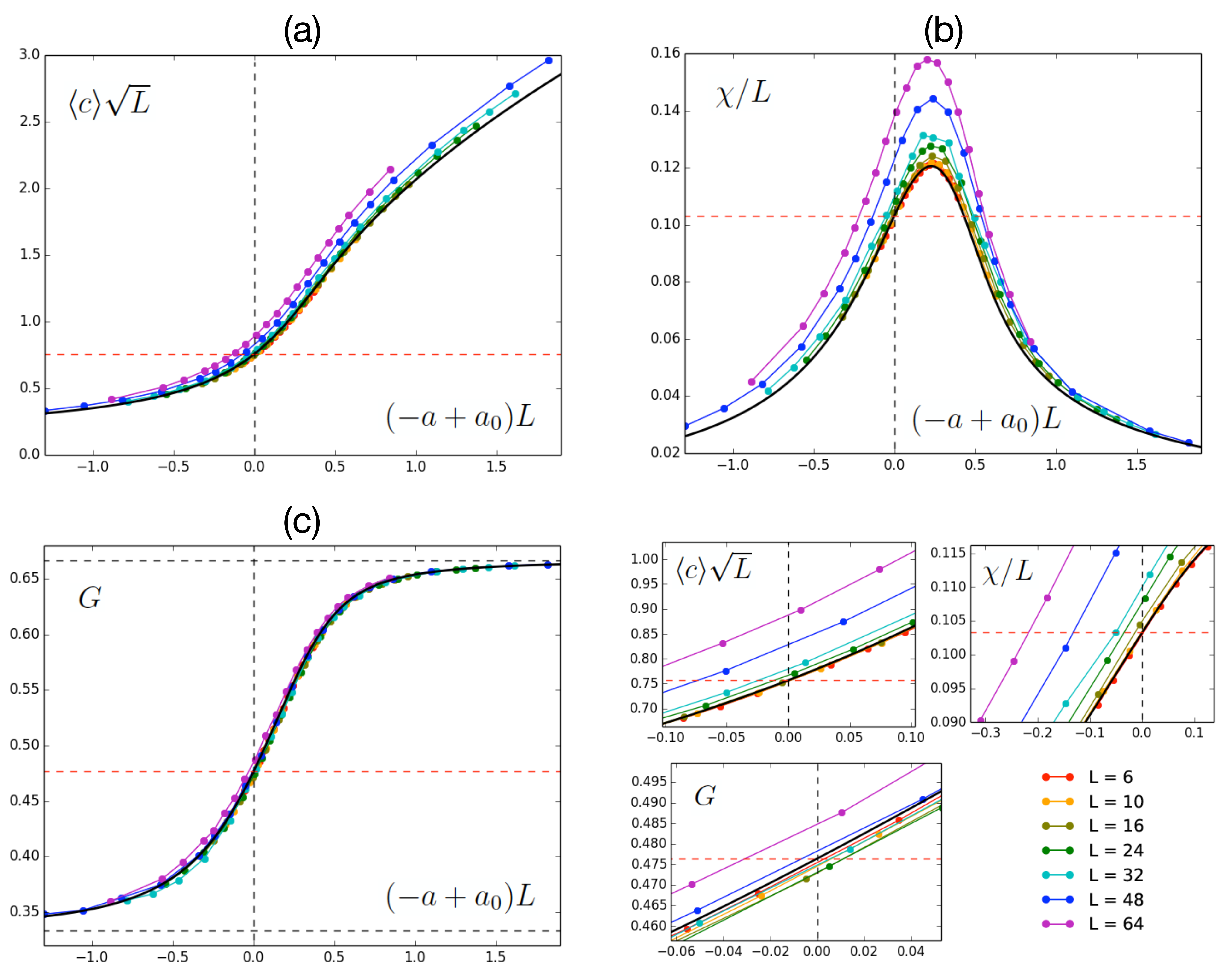}
\caption {Scaling plots of (a) mean order parameter $\langle c \rangle$, (b) susceptibility $\chi$ and (c) Binder cumulant $G$ for the ``ITT-5s" simulations at various sizes $L$, according to the weak-coupling critical theory. The theory predicts that these scaled quantities collapse onto the scaling functions (black lines); each scaling function takes the value (\ref{eq:F10})--(\ref{eq:GcWCCT}) (red dashed line) at a kind of critical point $a = a_0$ (vertical dashed line). The limiting Binder cumulant values $1/3$ for disorder and $2/3$ for order are marked by black dashed lines in (c), and a magnified view near $a = a_0$ is also shown. The simulation data are in good agreement with the theory at the smallest sizes $L = 6, 10, 16$ in Regime I, and they gradually deviate from the predicted collapse as size further increases.}
\label{fig:collapsesfiv}
\end{figure}

Next we compare the predicted FSS forms (\ref{eq:cscalingformWCCT})--(\ref{eq:GscalingformWCCT}) against numerical simulation. Apparently the WCCT exponent $\gamma/\nu = 1$ agrees with the susceptibility peak scaling $\chi_{\max}(L) \sim L^{\gamma/\nu}$ of the ITT simulations in Regime I, as shown in figure \ref{fig:susmax}. The detailed scaling forms of the mean order parameter, the susceptibility and the Binder cumulant can be further compared.  For each size in ``ITT-5s", we numerically evaluate the $N$-dependent critical point $a_0(N = L/\Delta x)$ defined in (\ref{eq:aoWCCT}), and plot in figure \ref{fig:collapsesfiv} the scaled quantities $\langle c \rangle \sqrt{L}$, $\chi/L$ and the Binder cumulant $G$ versus $[-a + a_0(N)] L$. The scaling functions in integral forms as given in \ref{app:scaling} are numerically evaluated and plotted as the black lines in figure \ref{fig:collapsesfiv}. The simulation data show good agreement with the theory at the smallest sizes $L= 6, ~10, ~16$ in Regime I, and display a deviation from the collapse as $L$ further increases.

 The WCCT also explains the $\Delta x$-independence of the mean-flow statistics such as the susceptibility peak value $\chi_{\max}$ and the Binder cumulant crossing value $G_{\rm c}$ in Regime I as observed in numerical simulation. Keeping the model parameters $\{ b, ~\mu, ~D \}$ fixed, at the same size $L$, two simulations done at different resolutions $\Delta x$ correspond to the same mean-flow dynamics (\ref{WCCT-eq1-2}) with a control parameter $a_{\rm R}$ left to be tuned, and thus the statistics at criticality will be identical. The only difference lies in the eddy-induced shift in the critical point: The positive quantity $-a_0(N)$ as given in (\ref{eq:aoWCCT}) includes contributions from more large-$k$ modes at a finer resolution, and thus the bare linear coefficient $-a$ has to shift to a larger positive value (i.e., the ordered side) to compensate for the extra fluctuations and to maintain the system at criticality. 
  
\begin{figure}
\centering
\includegraphics[width=14.0 cm]{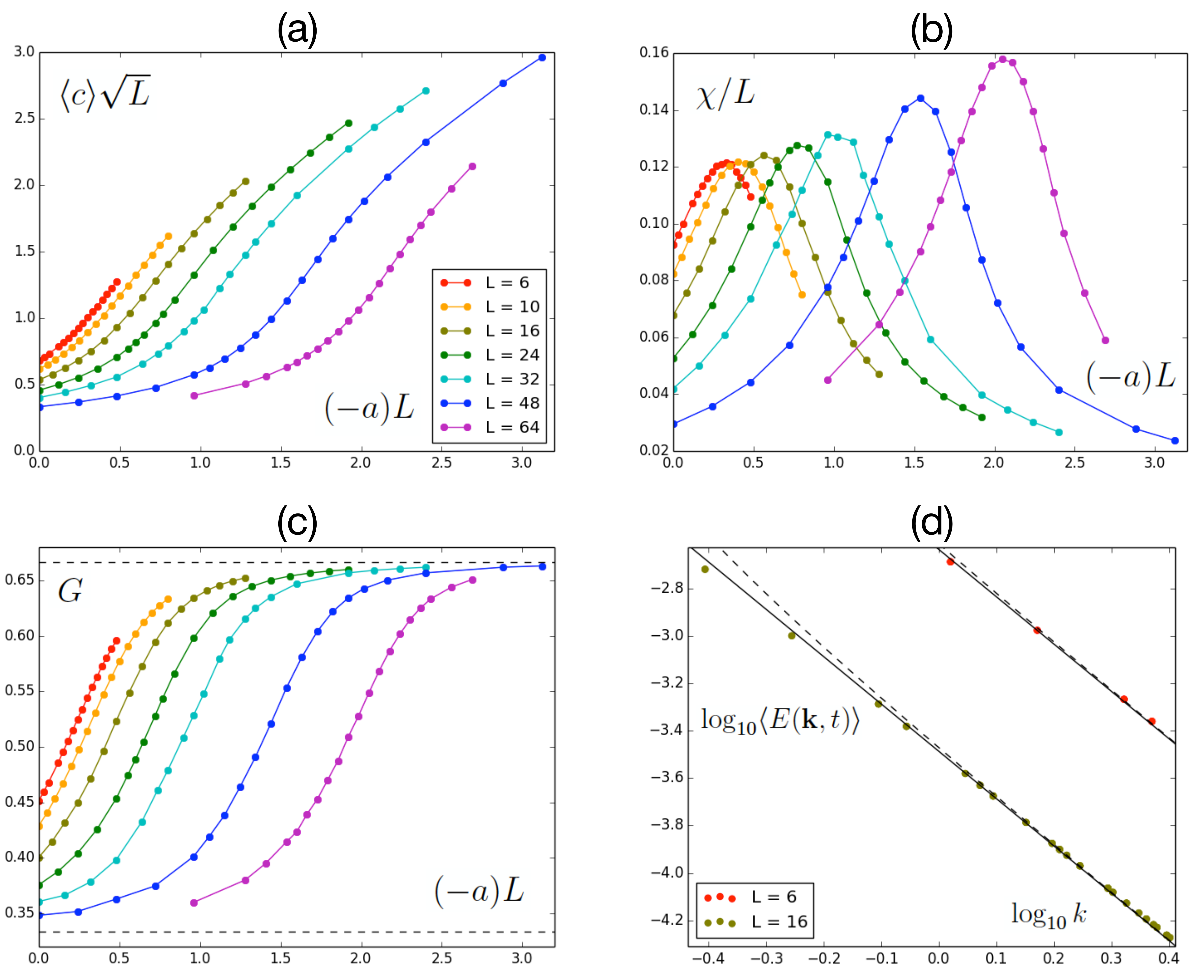}
\caption {Comparison between the ``ITT-5s" simulations and other candidate theories. (a)--(c) Scaling plots of the same data as figure \ref{fig:collapsesfiv} but according to mean field theory. The eddy-induced shift of the critical point $a_0$ that is considered in the weak-coupling critical theory is neglected in mean field theory; the horizontal non-collapse indicates that this eddy ($k>0$) effect on the mean flow ($k = 0$) is not negligible in Regime I. (d) The average power spectrum for $L = 6$ and $L = 16$ (both in Regime I) agrees with the Gaussian solution (solid line) better than the linear theory (dashed line). The data shown in figure \ref{fig:powerEk} are replotted here. The linear theory takes the bare linear coefficient $a = a_{\rm c}^{\rm sus.}(L)$, whereas $a = 0$ for the Gaussian fixed point. Numerical results show that the eddy dynamics in Regime I is not simply linear, and some nonlinear coupling works to effectively cancel out the linear term.}
\label{fig:mft-linear}
\end{figure}

Thus this simple WCCT is successful in quantitatively describing Regime I. The reader may further wonder whether a still simpler theory such as mean field theory or the linear theory is already sufficient, and what is the relation between them and the WCCT. Note that the WCCT includes physics beyond mean field theory (or, the so-called lowest-mode theory in the FSS theory first proposed by Br{\'e}zin and Zinn-Justin \cite{BZ1985,JLS2007}), though the exponents $\gamma = 1$, $\nu = 1$, $\beta = 1/2$, and $z = 1$ are mean-field. Mean field theory corresponds to neglecting all the eddies in (\ref{WCCT-eq1}), leading to the same result but with $a_0 = 0$. A scaling plot based on mean field theory shows a clear non-collapse of the phase transition curves in the horizontal direction (see figures \ref{fig:mft-linear}(a)--(c)). Mean field theory thus fails to capture the eddy-induced shift of the critical point which is not a small effect. Moreover, we stress that the Gaussian assumption for the eddies does not mean the absence of nonlinear effects in the $k > 0$ dynamics. In the linear theory, equation (\ref{eq2}) becomes
\begin{eqnarray}
 \frac{\partial \omega}{\partial t} &=& -a \omega + \mu \nabla^2 \omega + \eta({\bf r}, t),
\end{eqnarray}
which differs from the {\it critical} linear dynamics (\ref{WCCT-eq2}) by an additional linear term that is nonzero at the finite-size critical point $ a = a_{\rm c}^{\rm sus.}(L) < 0$. Naively one may expect the Gaussian assumption as a consequence of the linear theory for small systems: When sizes are small, the wavenumbers are all large enough to guarantee $|a| \ll \mu k^2 $, so the linear theory is approximately the Gaussian fixed point. Our data disagree with this naive expectation. Figure \ref{fig:mft-linear}(d) compares the measured average power spectrum with the Gaussian prediction (\ref{eq:power}) and with the linear prediction (derived in \ref{app:linear}) 
\begin{eqnarray}
\overline{ E({\bf k}, t) }_{\rm linear}  =     \frac{  D  }{(a + \mu k^2) L^2 } , 
\end{eqnarray}
where the bare parameters of each simulation are used. In the small-$k$ regime where the relation $|a| \ll \mu k^2 $ breaks down, the data points still lie closer to the Gaussian prediction than to the linear one. Thus at the finite-size critical point, some nonlinear coupling in the eddy dynamics is present to effectively cancel out the linear term.

In large enough systems such as Regimes II and III, the $k = 0$ mode and the $k > 0$ modes show the same set of universal exponents that corresponds to a specific RG fixed point; by contrast, the WCCT highlights the interesting feature of Regime I that when size is small enough to constrain the nonlinear effects severely, $k = 0$ and $k > 0$ show different sets of exponents, namely mean-field and Gaussian respectively. The Gaussian exponents are contrasted with mean-field ones in table \ref{table:mftvsgaussian}. All these listed mean-field exponents for $k = 0$ and the Gaussian exponent $\eta = 0$ for $k > 0$ are directly confirmed by our numerical measurements; we defer direct verification of the remaining Gaussian exponents for the $k > 0$ modes to future work.

\begin{table}
 \caption{ \label{table:mftvsgaussian}Exponents of mean field theory and those of the Gaussian fixed point in 2D. In Regime I mean-field exponents are observed for the $k = 0$ mode, whereas Gaussian statistics is found for $k >0$.}
 \begin{indented}
\item[]\begin{tabular}{@{} c  c  c   c  c  c }
\br
  & $\gamma/\nu$ &  $\beta/\nu$  &   $z$ & $1/\nu$ & $\eta$ \\
\mr
 Mean-Field & 1 & 1/2 & 1 & 1 & --- \\
 Gaussian  & 2 & 0 & 2 & 2 & 0 \\
\br
\end{tabular}
\end{indented}
 \end{table}

\section{Conclusions and discussion \label{sec:conclusions}}

We have performed numerical simulations of the 2D ITT equation on the square torus to study the random-to-flocking transition. The transition is observed to be continuous, which was also found in the perturbative dynamical RG analysis near 4D \cite{CTL2015}. We have further investigated FSS of various statistical quantities associated either with the homogeneous $k = 0$ mode or with the inhomogeneous $k > 0$ modes. A combination of numerical facts and analytical theories reveal a relatively complete picture of three FSS regimes. Regime I is an equilibrium regime shared by both the ITT model and the dipolar XY continuum model. Here the particle-level ferromagnetic alignment interactions lead to the cubic nonlinearity, but due to severe finite-size constraints, this nonlinearity only acts weakly on the dynamics as depicted in the WCCT. Regime II is also an equilibrium regime shared by the two continuum models, and it is the asymptotic scaling regime of the dipolar XY universality class. Regime III is the off-equilibrium regime of the ITT class where the advective nonlinearity further comes into play. 

The crossover from Regime II to Regime III reveals how activity affects the formation of long-range orientational order. From the perspective of hydrodynamics, activity appears as the additional advection term compared to the equilibrium ferromagnetic systems. In Regime II vortices are excited by the equilibrium thermal noise  \cite{VTMFF2014}, whereas in Regime III, advection comes in to drive the vortex dynamics, producing non-equilibrium critical fluctuations of an active fluid. Numerical simulation demonstrates that activity helps to stabilize long-range order and facilitates information transfer faster than diffusion. That advection accelerates ordering processes was also found in coarsening via vortex merger events \cite{RP2020}. The underlying non-equilibrium vortex physics remains to be understood.

Through quantitative comparison with numerical simulations, we established a novel theoretical approach of reduced hydrodynamics to describe FSS for small systems at criticality. The WCCT captures accurately the statistics of the $k=0$ mode in Regime I, from which the observed but unexpected mean-field scaling is derived.
%and more work is needed to extend it to describe the first crossover.
Compared to our approach, the conventional theoretical approach, namely the RG flow equations was shown to be incomplete: In Regime I with a dominant mean flow, RG overlooks the difference in scaling properties between the $k = 0$ and $k > 0$ modes and thus only partially describes the large-scale behavior. As vortex excitations disrupt the mean flow near the first crossover, coupling between the $k=0$ and $k > 0$ modes is expected to grow and eventually invalidate the mean-field description. We leave a detailed discussion of this process to future work.

%The RG-based FSS theory \cite{BZ1985,JLS2007} deals with the homogeneous mode, which is formally similar to the WCCT, but the two theories differ in procedures of tracing out the inhomogeneous modes and have different regimes of validity. In the WCCT these modes are simply placed at the Gaussian fixed point, whereas in the RG-based theory, they are integrated out by the perturbative RG procedure in an $\epsilon$-expansion. The RG-based theory works for large critical systems near the upper-critical dimension, whereas the WCCT has been shown here to work for small critical systems of the 2D ITT model and 2D dipolar XY continuum model, even though $d = 2$ is far below the upper-critical dimension $d^{*} = 4$.

Finally, we would like to emphasize that the small-size regime, i.e., Regime I as described by the WCCT, has the interesting feature that depending on whether the measurement is done on the $k = 0$ mode or on the $k > 0$ modes, the resulting exponents can either be mean-field or Gaussian. This regime also exhibits the distinct velocity patterns that are relatively ordered with almost no vortex. It emerges here as system size constrains the ferromagnetic nonlinear effects in the incompressible fluids, and the other nonlinearity does not come into play. Such a regime may also appear in other incompressible systems where ferromagnetic alignment interactions are present, and in other dimensions such as 3D. Following \cite{CCGGP2021}, the incompressibility condition may be relaxed: It may also be possible to find this regime in compressible systems such as the Vicsek model at the onset of order, when inhomogeneous density structures do not appear due to limited system size; this regime may be further related to certain scale-free behavior in natural flocks.

\ack
We thank Beno\^{\i}t Mahault, Aurelio Patelli, Xia-qing Shi, and Leiming Chen for helpful discussions. The work is supported in part by the National Science Foundation of China (NSFC) under Grant No. 11635002, and by the Research Grants Council of the Hong Kong Special Administrative Region (HKSAR) under Grant No. 12304020.
We acknowledge the computational support from the Beijing Computational Science Research Center (CSRC) and Special Program for Applied Research on Super Computation of the NSFC-Guangdong Joint Fund (the second phase). WQ acknowledges financial support by Basic Science Research Program through the National Research Foundation of Korea (NRF) funded by the Ministry of Education (20211060) and Korea government (MSIT) (No.2020R1A5A1016518).

\appendix

\section{Supplementary transition curves \label{app:trans}} 
\begin{figure}
\centering
\includegraphics[width=17.0 cm]{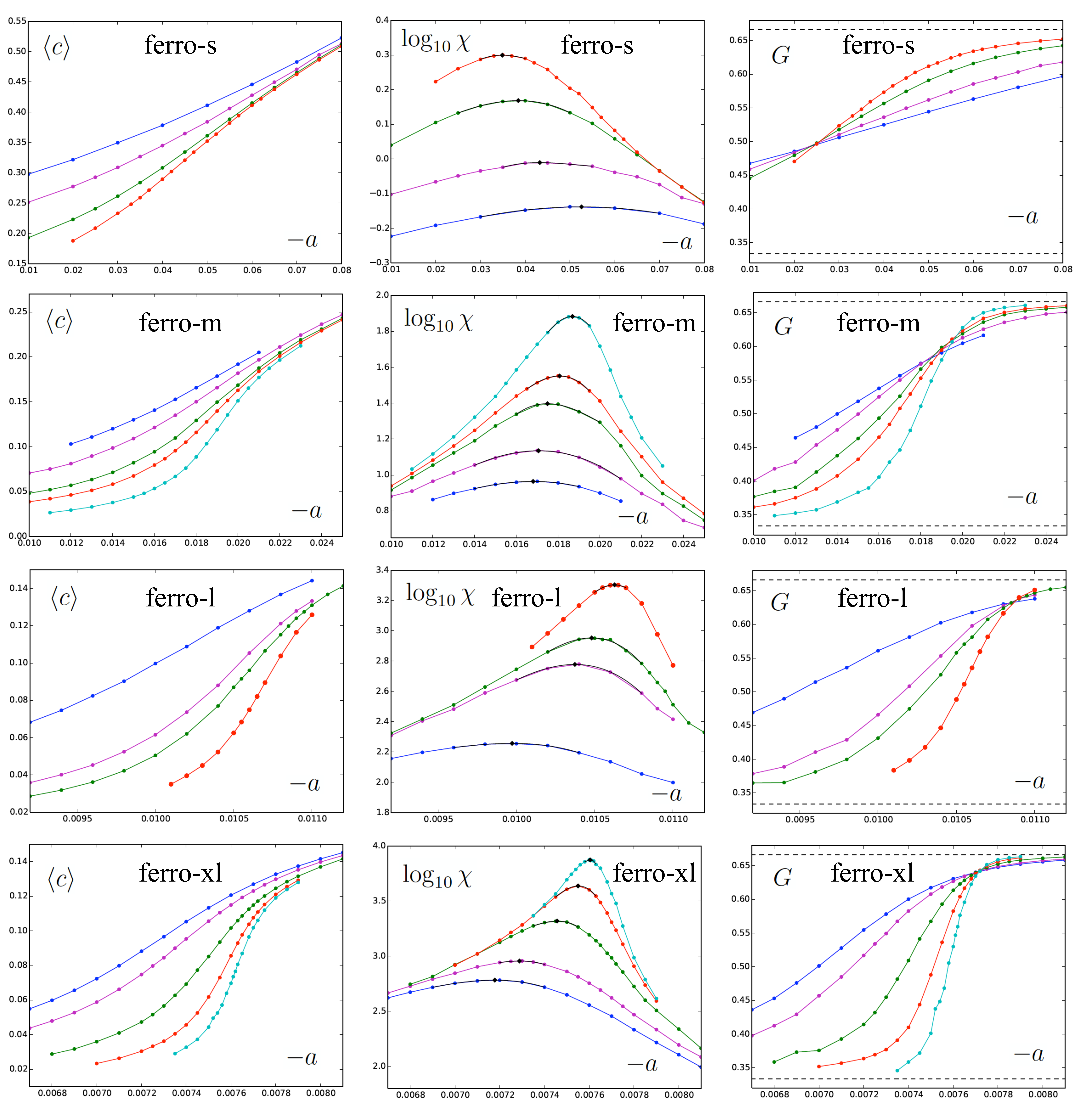}
\caption {Same as figure \ref{fig:trans-five} but for various ``ferro" simulations. The ``ferro-l1" and ``ferro-l2" data are plotted in the same figures as denoted by ``ferro-l", because these simulations are performed at the same spatial resolution $\Delta x$, though with different time steps $\Delta t$. We have numerically verified for several cases that for $\Delta t$ small enough to make the simulation stable and well-behaved, changing $\Delta t$ does not change $\langle c \rangle$, $\chi$ and $G$. System sizes: ``ferro-s" with $L = 6, ~8, ~12, ~16$, ``ferro-m" with $L = 60, ~80, ~120, ~150, ~240$, ``ferro-l1" with $L = 400, ~800, ~1000$, ``ferro-l2" with $L = 1600$, and ``ferro-xl" with $L = 800, ~1000, ~1600, ~2400, ~3200$. }  
\label{fig:trans-ferro}
\end{figure}

\begin{figure}
\centering
\includegraphics[width=17.0 cm]{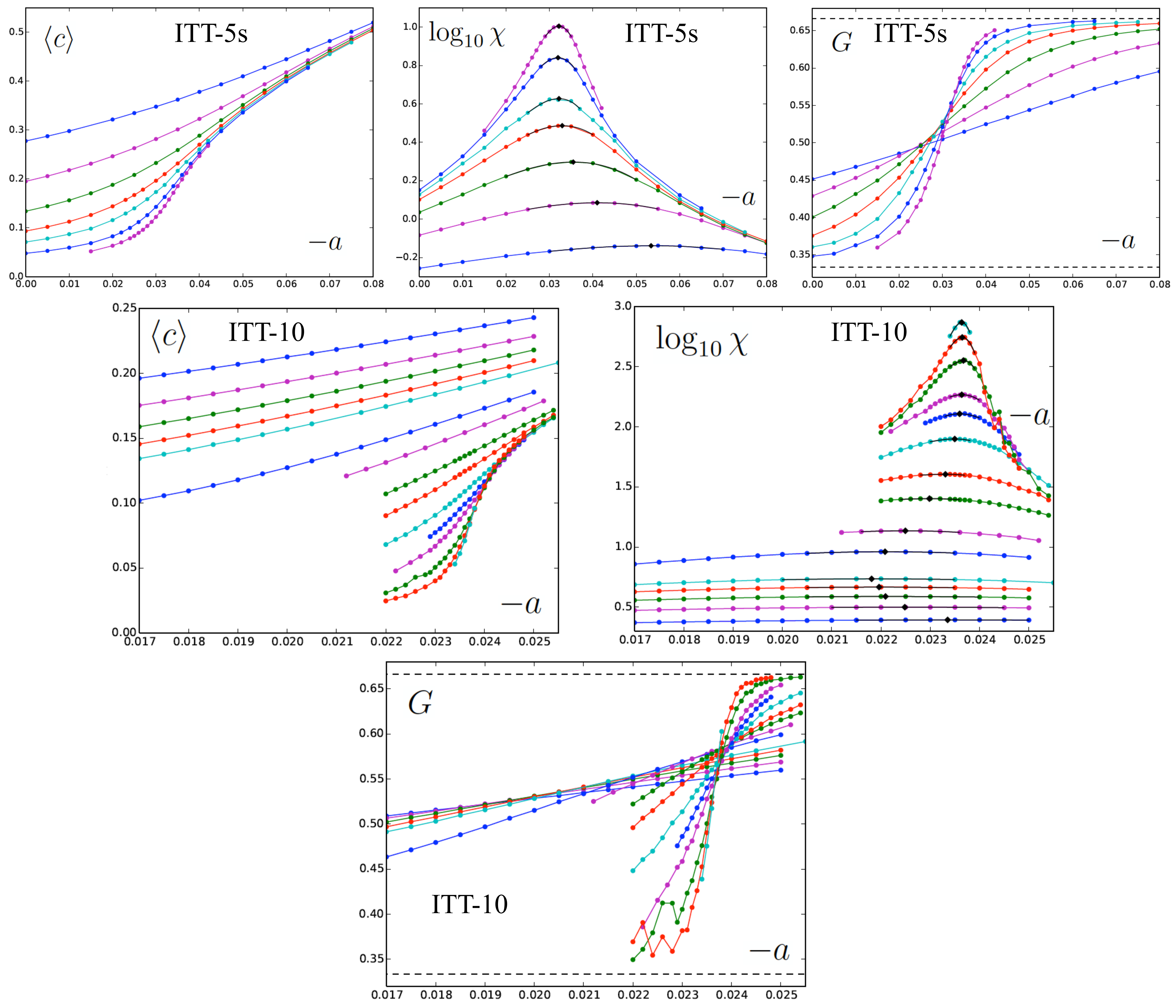}
\caption {Same as figure \ref{fig:trans-five} but for the other ITT simulations. System sizes: ``ITT-5s" with $L = 6, ~10, ~ 16, ~24, ~32, ~ 48, ~64$, and ``ITT-10" with $L = 20, ~25, ~30, ~35, ~40, ~60, ~80, ~120, ~160, ~240, ~320, ~400, ~600, ~800, ~960$.  }  
\label{fig:trans-sfiv-tenss}
\end{figure}

The phase transition curves of the mean order parameter $\langle c \rangle$, the susceptibility $\chi$ and the Binder cumulant $G$ for the ``ferro" simulations (when the advection coefficient $\lambda = 0$) at various resolutions are shown in figure \ref{fig:trans-ferro}. Those for ``ITT-5s" and ``ITT-10" are presented in figure \ref{fig:trans-sfiv-tenss}. The statistical error can be estimated from the local smoothness of the transition curves.

\section{Susceptibility peak value and Binder cumulant crossing value at different resolutions  \label{app:diffdx}}
\begin{figure}
\centering
\includegraphics[width=15.0 cm]{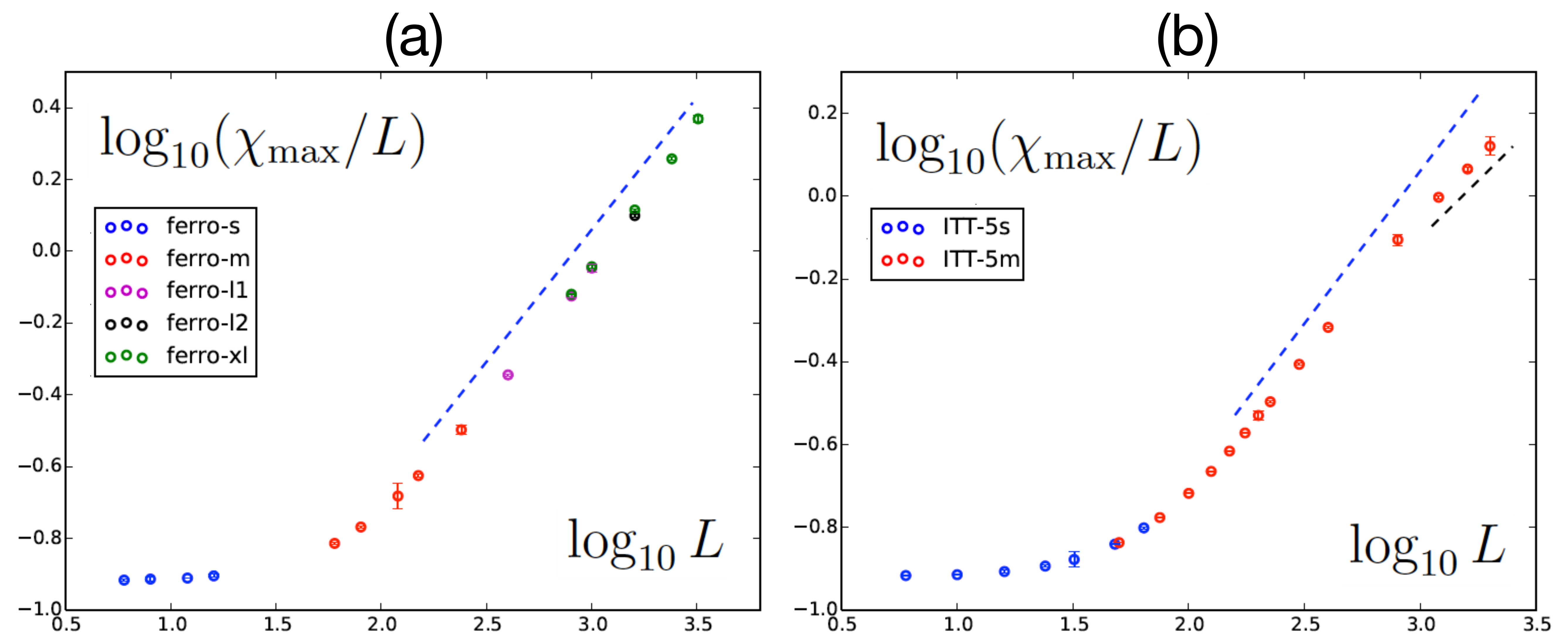}
\caption {Divergence of the susceptibility peak value $\chi_{\max}$ with system size $L$ for (a) ``ferro" and (b) ``ITT-5" simulations, as measured at different spatial resolutions $\Delta x$. The black dashed line marks the $\chi_{\max} \sim L^{1.55}$ power law, and the blue dashed line marks $\chi_{\max} \sim L^{1.737}$ of the dipolar XY class \cite{MC2010}. The susceptibility peak value is found to be insensitive to $\Delta x$ in Regimes I and II.}  
\label{fig:susmaxdiffdx}
\end{figure}

\begin{figure}
\centering
\includegraphics[width=15.0 cm]{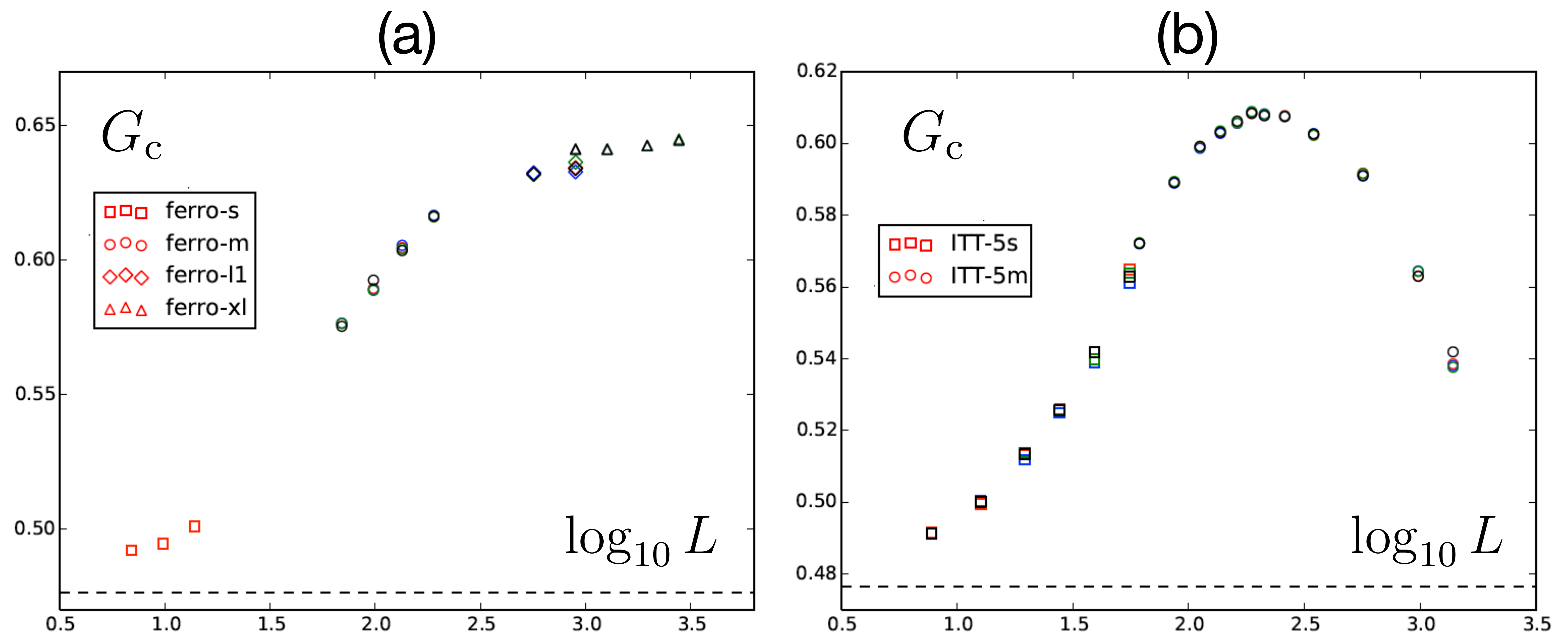}
\caption {Binder cumulant crossing value $G_{\rm c}$ versus $\log_{10}L$ for (a) ``ferro" and (b) ``ITT-5" simulations, as measured at different spatial resolutions $\Delta x$. Different $\Delta x$ corresponds to different symbol shape, and different colors represent different fitting methods as in figure \ref{fig:mucL}. The dashed line marks the value $1-\pi/6$ as predicted by the weak-coupling critical theory. The Binder cumulant crossing value seems insensitive to $\Delta x$ in Regimes I and II. }  
\label{fig:GcLdiffdx}
\end{figure}

The susceptibility peak value $\chi_{\max}(L; \Delta x)$ and the Binder cumulant crossing value $G_{\rm c}(L; \Delta x)$ are found to be insensitive to spatial resolution $\Delta x$ in Regimes I and II. Figure \ref{fig:susmaxdiffdx} compares the susceptibility peak value for ``ferro" and ``ITT-5" at different resolutions. In the absence of the advection term, the ``ferro" simulations only have Regimes I and II but no Regime III. Except for the three largest-$L$ points of ``ITT-5" that lie outside Regimes I and II, the data points of different $\Delta x$ fall onto a single curve within statistical uncertainty, showing that the susceptibility peak value is insensitive to $\Delta x$. The Binder cumulant crossing value at different resolutions is compared in figure \ref{fig:GcLdiffdx}. We again find that, except for the $\log_{10}L \gtrsim 2.3$ part of ``ITT-5" that deviates from the ``ferro" behavior and is outside Regimes I and II, the other data points of different $\Delta x$ almost fall onto a single curve. Thus the Binder cumulant crossing value also seems insensitive to $\Delta x$ in Regimes I and II.

\section{Size-independent power spectrum and abnormal tail \label{app:pow-diffL}}

\begin{figure}
\centering
\includegraphics[width=15.0 cm]{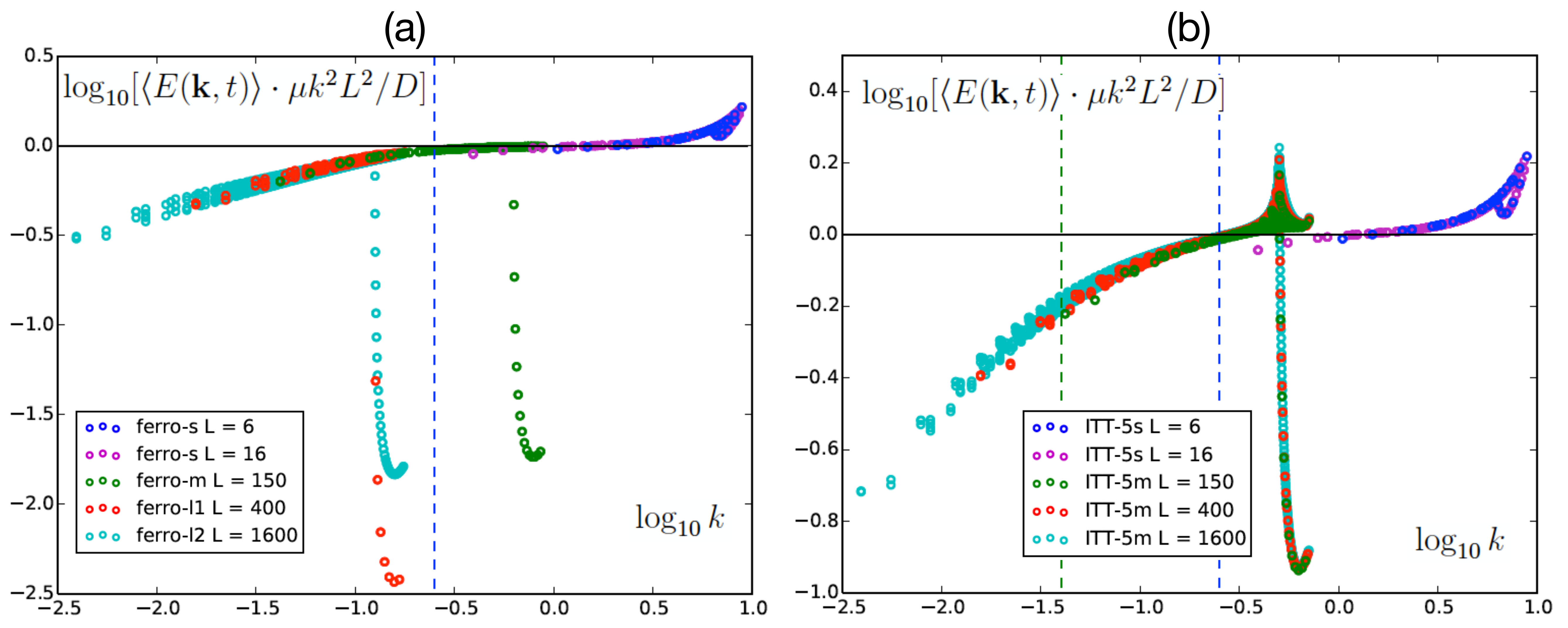}
\caption {Same as figure \ref{fig:powerEk} but now different $\{ \Delta x, L\}$ are distinguished for (a) ``ferro" and (b) ``ITT-5" simulations. Ignoring the abnormal tails at the largest wavenumbers in each simulation, we find that the scaled average power spectrum $\langle E({\bf k}, t) \rangle L^2$ is independent of size $L$ and resolution $\Delta x$.  }  
\label{fig:pow-diffL}
\end{figure}

When the model parameters $\{D, ~\mu, ~\lambda, ~b \}$ are fixed, the scaled average power spectrum $\langle E({\bf k}, t) \rangle L^2$, measured at the susceptibility peak location $a_{\rm c}^{\rm sus.}(L;\Delta x)$, is found to be independent of system size $L$ and spatial resolution $\Delta x$, as shown in figure \ref{fig:pow-diffL}. Here for either ``ferro" or ``ITT-5", the data points of different $\{ \Delta x, L\}$ fall onto the same curve.

\begin{figure}
\centering
\includegraphics[width=9.0 cm]{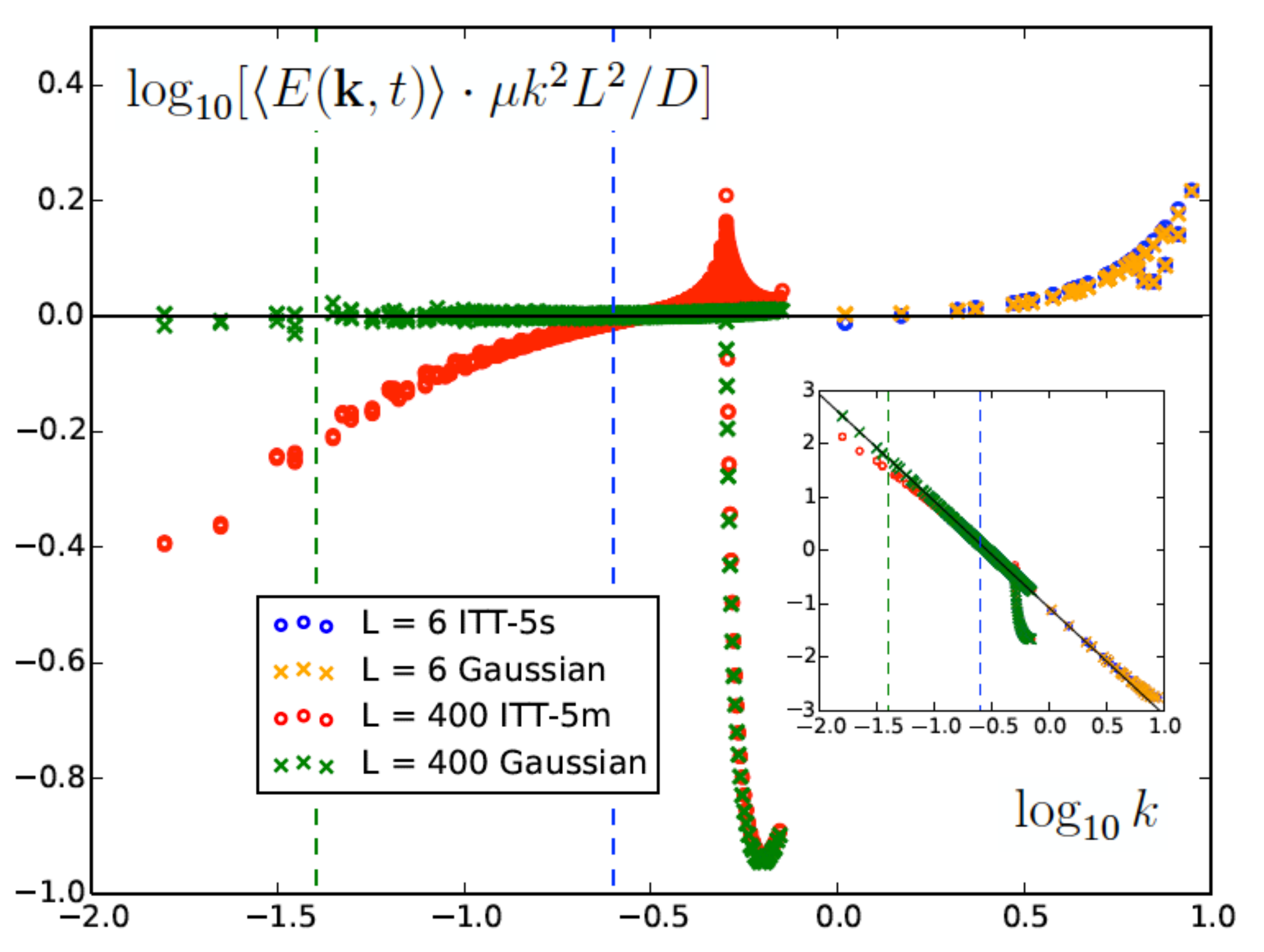}
\caption {Same as figure \ref{fig:powerEk} but here two ``ITT-5" simulations are compared with numerical simulation of the critical linear model (or, the Gaussian fixed point). Two ITT simulations at different $\{\Delta x, ~L\}$ as shown in figure \ref{fig:pow-diffL}(b) are compared against the corresponding simulations of the critical linear model. The data show that the abnormal tail at the largest wavenumbers in each simulation is only a numerical artifact. The $k^2$ factor in the scaled average power spectrum exaggerates the abnormal tail; the original size-independent power spectrum $\langle  E({\bf k}, t)\rangle L^2$ is plotted against wavenumber $k$ in $\log_{10}-\log_{10}$ scales in the inset. }  
\label{fig:pow-tail}
\end{figure}

The data points deviate from this curve at the largest wavenumbers in each simulation; this abnormal tail is a numerical artifact due to finite spatial resolution in the pseudo-spectral scheme. To show this, we simulate the critical linear model (\ref{WCCT-eq2}) using the same numerical scheme as ITT but now with the parameters $a = 0$, $b = 0$ and $\lambda = 0$. The other parameters $\{ \mu, ~ D \}$, spatial resolution $\Delta x$, time step $\Delta t$ and system size $L$ are kept the same as the two ``ITT-5" simulations shown in figure \ref{fig:pow-tail}. The measured power spectrum of the critical linear model should agree with the analytical solution (\ref{eq:power}), if there is no numerical artifact. However, similar abnormal tails appear (see figure \ref{fig:pow-tail}), confirming that this behavior is a numerical artifact. We have further checked for the critical linear simulation, when $\Delta x$ is decreased, the abnormal tail appears at a larger $k$, and is thus absent in the finite $k$-range in the continuum limit. As has been checked for some of the ITT simulations, using anti-aliasing with 1/2 rule for cubic nonlinearity only truncates the tail, but does not change the values of $\langle { E({\bf k}, t) } \rangle$ at the retained modes, and also does not eliminate this artifact. 

\section{Autocorrelation collapse and autocorrelation time at different resolutions \label{app:tau-related}}

\begin{figure}
\centering
\includegraphics[width=10.0 cm]{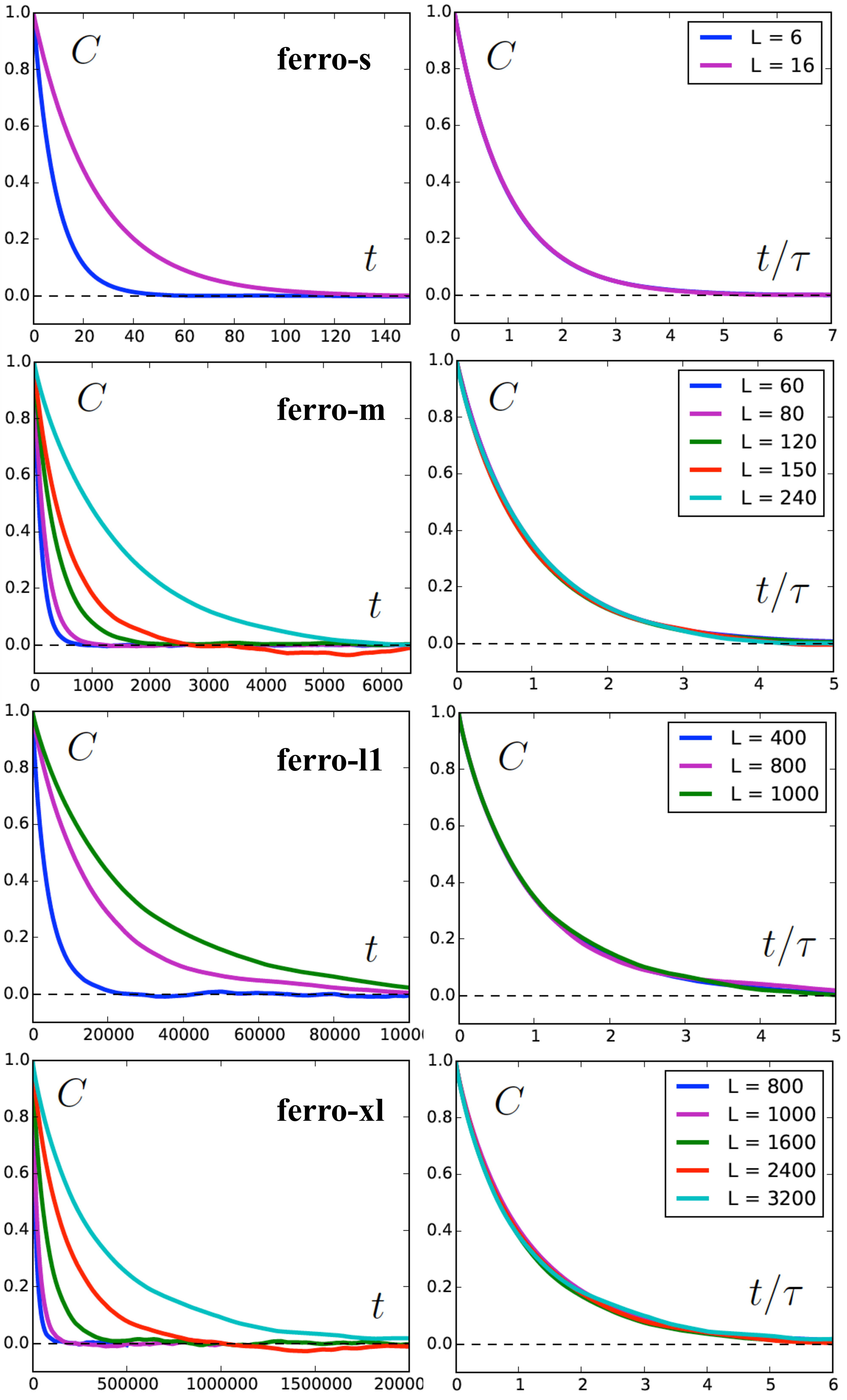}
\caption {Scaling collapse of autocorrelation function for various ``ferro" simulations at different sizes $L$. The original autocorrelation function $C(t)$ is shown on the left, and on the right, autocorrelation is plotted against the rescaled lag time $t/\tau$ using the measured values of autocorrelation time $\tau(L)$. The collapse confirms the accuracy of the measured $\tau$-values for each autocorrelation function.}  
\label{fig:ferro-corr}
\end{figure}

\begin{figure}
\centering
\includegraphics[width=11.0 cm]{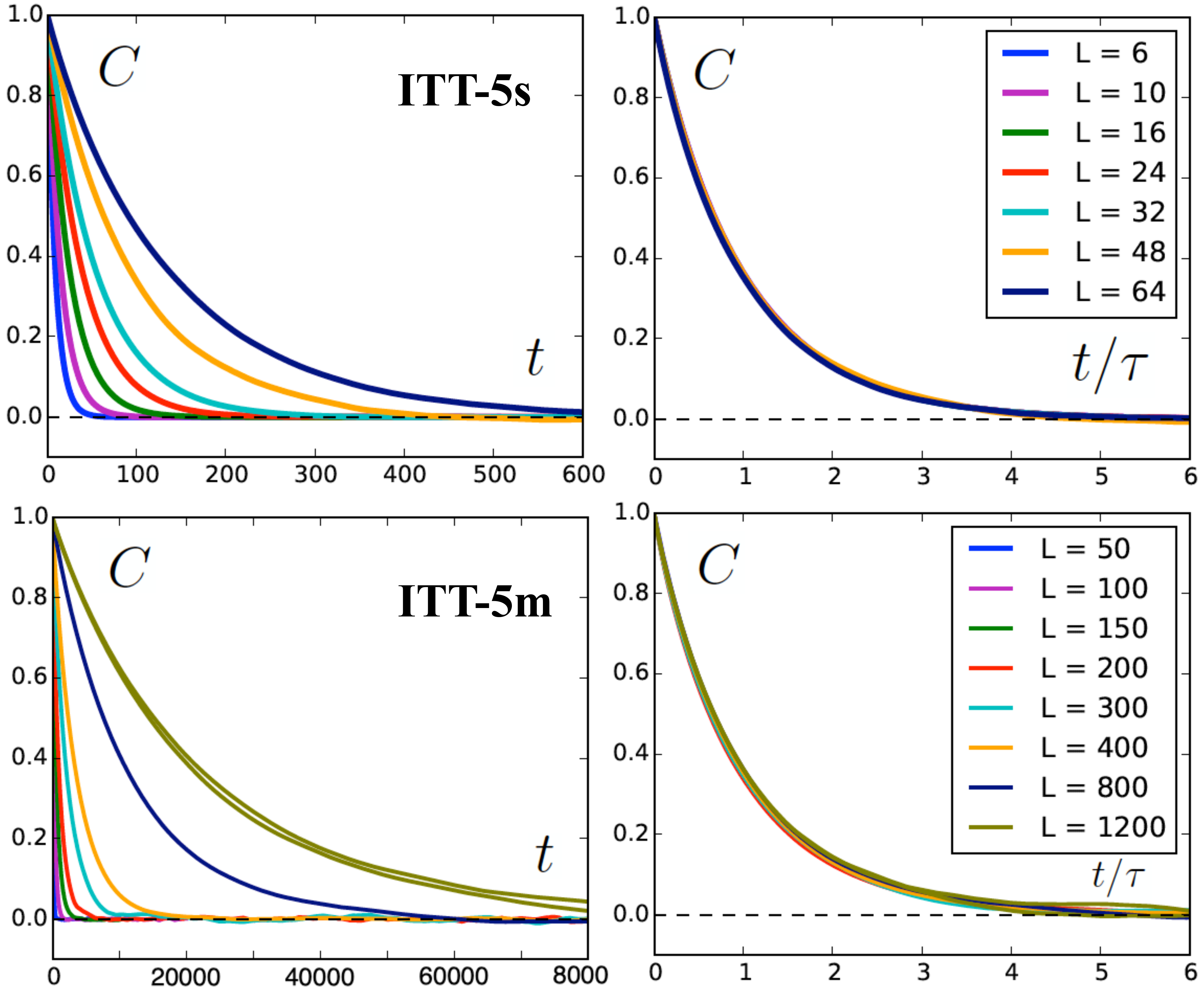}
\caption {Same as figure \ref{fig:ferro-corr} but for the ``ITT-5" simulations. The $L= 1200$ case of ``ITT-5m" consists of two independent measurements with the same sample size; each measurement yields a $\tau$-value in figure \ref{fig:tau}, and the scatter between the two provides a rough estimate of error. Again the collapse on the right shows that the measured $\tau$-values are accurate.}  
\label{fig:ITT5-corr}
\end{figure}

\begin{figure}
\centering
\includegraphics[width=15.0 cm]{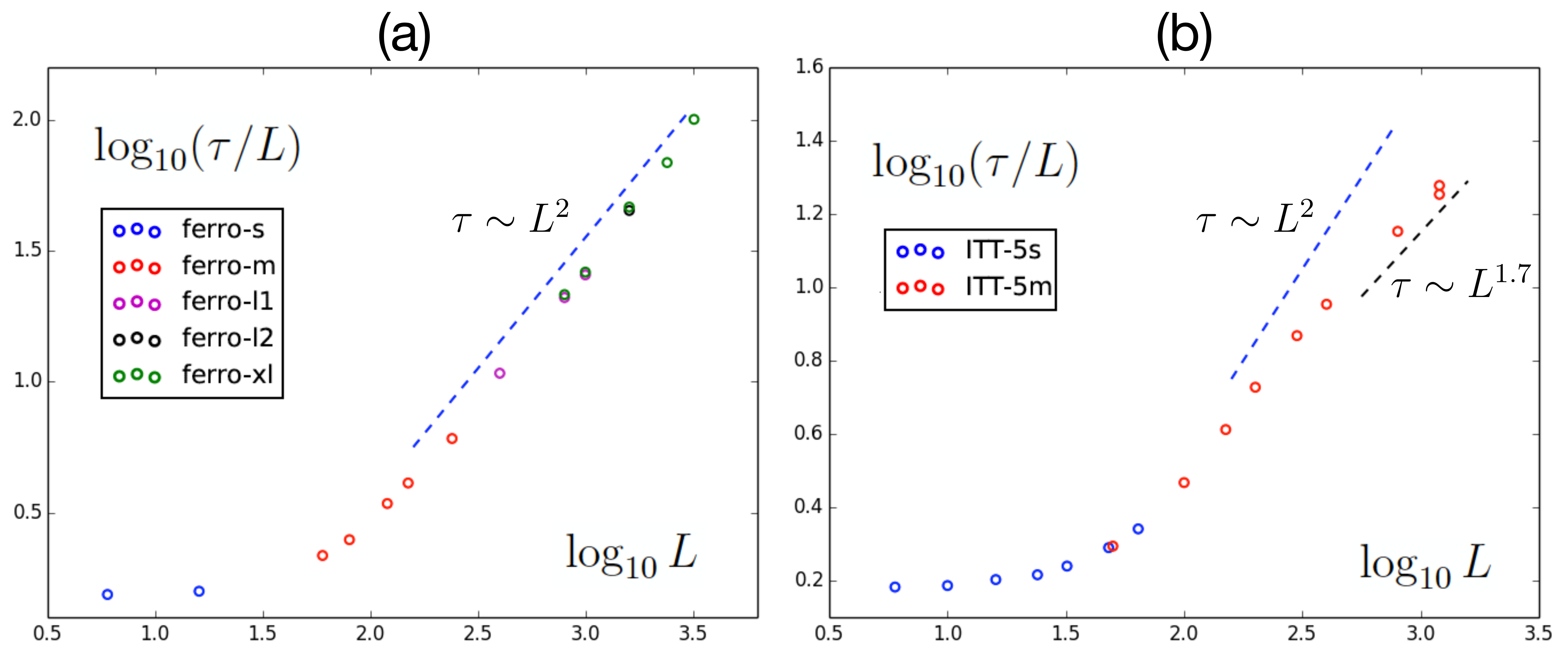}
\caption {Divergence of autocorrelation time $\tau$ with system size $L$, for (a) ``ferro" and (b) ``ITT-5" simulations at different spatial resolutions $\Delta x$ . The autocorrelation time is found to be insensitive to $\Delta x$ in Regimes I and II.}  
\label{fig:tau-diffdx}
\end{figure}

The autocorrelation time $\tau$ is obtained by fitting the autocorrelation function $C(t) \sim \rme^{-t/\tau}$ for large lag time $t$. The fitting range should be chosen properly: The decay at shorter lag time is faster than exponential, and the autocorrelation data become too noisy to fit when lag time is too large. 
Figures \ref{fig:ferro-corr} and \ref{fig:ITT5-corr} compare the original autocorrelation function $C(t)$ with the autocorrelation as a function of the rescaled time $C(t/\tau)$, for different sets of simulations. The collapse of $C(t/\tau)$ for different system sizes confirms the accuracy of our measured $\tau(L)$ values. Figure \ref{fig:tau-diffdx} further shows that the autocorrelation time $\tau(L)$ is insensitive to spatial resolution $\Delta x$ in Regimes I and II. Except for the $\log_{10}L \gtrsim 2.6$ part of ``ITT-5" that is outside Regimes I and II, the other data points of different $\Delta x$ approximately fall onto the same curve. 

\section{Scaling argument that Regime III should follow Regime II \label{app:scaling_argument}}

In Regime II of the dipolar XY class, if we ignore the diagrammatic corrections and only perform rescaling in the RG procedure to go to larger scales, the advection coefficient will evolve according to
\begin{eqnarray}
\frac{\rmd \lambda}{ \rmd \ell} = (\chi -1+z )\lambda,
\end{eqnarray} 
 where $\ell$ is the RG flow parameter, and the dynamic exponent $z$ and the ``roughness" exponent $\chi$ are defined in rescaling ${\bf r} \rightarrow \rme^\ell {\bf r}$, $t \rightarrow \rme^{z \ell} t$ and ${\bf v} \rightarrow \rme^{\chi \ell} {\bf v}$. The sign of $(\chi -1 + z)$ is crucial in determining whether the advection coefficient will grow or decay exponentially. Our numerical measurements give the exponent $z \approx 2$, and without carefully estimating the uncertainty, we can use the safe bound $z > 1.7$ for Regime II. The ``roughness" exponent is related to the scaling of power spectrum by $\chi = (2 -d - \eta)/2$, and is also related to the scaling of the susceptibility peak value by $\chi  = (-d + \gamma/\nu)/2$. Our estimate $\eta = 0.27(8)$ yields $\chi = - 0.14(4)$, and the estimate $\gamma/\nu = 1.737(1)$ obtained from Monte Carlo simulations of the dipolar XY model \cite{MC2010} gives $\chi = -0.1315(5)$; both show that $\chi > -0.2$. Therefore, the quantity $\chi -1 +z > -0.2 -1 + 1.7 = 0.5 > 0$. The advection coefficient grows exponentially by power counting in Regime II, which supports that advection should come into play at larger sizes to bring the system into Regime III.

\section{Linear theory of eddies \label{app:linear}} 

We derive the power spectrum and other steady-state statistical properties of the linear model of eddies, as governed by
\begin{eqnarray}
  \frac{\partial \omega}{\partial t} &=&  -a \omega + \mu \nabla^2 \omega + \eta({\bf r}, t), \label{eqn:linear1}
 \end{eqnarray}
 where the vorticity noise $\eta \equiv \hat{\bf z}\cdot ( \nabla \times {\bf  f})$. The critical linear model, or the Gaussian fixed point, corresponds to the case with $a = 0$.

We introduce the Fourier transform and its inverse for the vorticity field (the frequency is denoted by $\Omega$ to not be mixed with the vorticity $\omega$),
 \begin{eqnarray}
 \omega({\bf k}, \Omega) &=& \int \rmd^2 {\bf r} \rmd t ~\omega({\bf r}, t) \rme^{\rmi(\Omega t - {\bf k} \cdot {\bf r})}, \label{eq:omegaFT}\\
 \omega({\bf r}, t ) &=& \int \frac{\rmd^2 {\bf k}}{(2 \pi)^2} \int \frac{\rmd \Omega}{2 \pi}~ {\omega}({\bf k}, \Omega) \rme^{-\rmi(\Omega t - {\bf k} \cdot {\bf r})}, \label{eq:omegaIFT}
 \end{eqnarray} 
 and likewise define ${\psi}({\bf k}, \Omega)$ for streamfunction, ${\bf v}({\bf k}, \Omega)$ for velocity, ${\eta}({\bf k}, \Omega)$ and ${\bf f}({\bf k}, \Omega)$ for noise.
 Equation (\ref{eqn:linear1}) yields
 \begin{eqnarray}
  \omega({\bf k}, \Omega) = \frac{ \eta({\bf k}, \Omega)}{a + \mu k^2 - \rmi \Omega},    \label{eq:omegaeta}
 \end{eqnarray}
and the vorticity-streamfunction relation (\ref{eq-omegapsi}) gives 
 \begin{eqnarray}
 \omega({\bf k}, \Omega) = - k^2  \psi({\bf k}, \Omega). \label{eq:omegapsi}
 \end{eqnarray}

To directly compare with numerical simulation, the discrete version has to be used instead. For example, (\ref{eq:omegaFT}) and (\ref{eq:omegaIFT}) become
 \begin{eqnarray}
 \omega({\bf{k}}, \Omega) &=&  \sum\limits_{i_{\rm t} = -N_{\rm t}/2}^{N_{\rm t}/2-1}~ \sum\limits_{i_{\rm x}, i_{\rm y} =-N/2}^{N/2-1}  \omega({\bf{r}}, t) ~\rme^{\rmi \frac{2 \pi}{N_{\rm t}} (n_{\rm t} i_{\rm t})-\rmi \frac{2 \pi}{N} (n_{\rm x} i_{\rm x} + n_{\rm y} i_{\rm y}) } \nonumber\\
 &\equiv& \sum\limits_{t} \sum\limits_{\bf r}  \omega({\bf{r}}, t) e^{\rmi \Omega t -\rmi {\bf k} \cdot {\bf r} } ,  \\
\omega({\bf{r}}, t ) &=& \frac{1}{ N^2 N_{\rm t}} \sum\limits_{n_{\rm t} = -N_{\rm t}/2}^{N_{\rm t}/2-1}~{\sideset{}{'}\sum\limits_{n_{\rm x}, n_{\rm y} = -N/2}^{N/2 -1}} \omega({\bf{k}}, \Omega) ~\rme^{-\rmi \frac{2 \pi}{N_{\rm t}} (n_{\rm t} i_{\rm t})+\rmi \frac{2 \pi}{N} (n_{\rm x} i_{\rm x} + n_{\rm y} i_{\rm y}) } \nonumber\\
&\equiv& \frac{1}{ N^2 N_{\rm t}}  \sum\limits_{\Omega} {\sideset{}{'}\sum\limits_{\bf k}}  \omega({\bf{k}}, \Omega) \rme^{-\rmi \Omega t +\rmi {\bf k} \cdot {\bf r} } , 
\end{eqnarray} 
where the prime on the summation means to exclude the wavenumber $(n_{\rm x}, n_{\rm y}) = (0, 0)$. Here ${\bf{r}} = \Delta x (i_{\rm x}, ~i_{\rm y} )$, $t = \Delta t \times i_{\rm t}$, ${\bf{k}} = \frac{2 \pi}{L} (n_{\rm x} , ~n_{\rm y} )$, $\Omega = \frac{2 \pi}{T} n_{\rm t}$, and $T$ is the measurement time. Note that $N= L/\Delta x$ and $N_{\rm t} = T/\Delta t$ are taken as even integers as in numerical simulations.

The vorticity noise correlation
 \begin{eqnarray}
 \overline{ {\eta}({\bf k}, \Omega) \cdot {\eta}^{*}({\bf k'}, \Omega')} &=& \overline{ [k_{\rm x} {f}_{\rm y}({\bf k}, \Omega) - k_{\rm y} {f}_{\rm x}({\bf k}, \Omega)] \cdot [k_{\rm x}' {f}_{\rm y}^{*}({\bf k'}, \Omega') - k_{\rm y}' {f}_{\rm x}^{*}({\bf k'}, \Omega')] } \nonumber\\
 &=& k^2 \frac{ 2 D ~N^2 N_{\rm t}}{(\Delta x)^2 \Delta t}  \delta_{n_{\rm t},n_{\rm t}'} \delta_{n_{\rm x}, n_{\rm x}'} \delta_{n_{\rm y}, n_{\rm y}'} \nonumber\\
 &=& k^2 \frac{ 2 D ~N^2 N_{\rm t}}{(\Delta x)^2 \Delta t}  \delta_{\Omega \Omega'} \delta_{{\bf k} {\bf k}'} ,  \label{eq:etacor}
 \end{eqnarray}
 where we have used the original noise correlation 
 \begin{eqnarray}
 \overline{ {f}_\alpha ({\bf k}, \Omega) {f}^{*}_\beta ({\bf k'}, \Omega') } = \frac{ 2 D ~N^2 N_{\rm t}}{(\Delta x)^2 \Delta t} \delta_{\alpha \beta}  \delta_{n_{\rm t},n_{\rm t}'} \delta_{n_{ x}, n_{\rm x}'} \delta_{n_{\rm y}, n_{\rm y}'}
 \end{eqnarray}
 which can be straightforwardly derived from the discrete version of (\ref{eq:noisef}). 
 Using (\ref{eq:omegaeta}), (\ref{eq:omegapsi}) and (\ref{eq:etacor}), we obtain the streamfunction correlation
 \begin{eqnarray}
 \overline{ {\psi}({\bf k}, \Omega) {\psi}^{*}({\bf k'}, \Omega')  } &=& \frac{1}{k^2} \frac{1}{(a + \mu k^2)^2 + \Omega^2} \frac{2D ~N^2 N_{\rm t}}{(\Delta x)^2 ~\Delta t} \delta_{\Omega \Omega'} \delta_{{\bf k} {\bf k}'}.  \label{eq:psicor}
 \end{eqnarray}
The velocity correlation for $k> 0$ and $k' > 0$ 
\begin{eqnarray}
 \overline{ {\bf v}({\bf k}, \Omega)\cdot {\bf v}^{*}({\bf k'}, \Omega') } &=&  (\hat{\bf z} \times {\bf k}) \cdot (\hat{\bf z} \times {\bf k'})~\overline{ {\psi}({\bf k}, \Omega)\cdot {\psi}^{*}({\bf k'}, \Omega')} \nonumber\\
  &=&  \frac{1}{(a + \mu k^2)^2 + \Omega^2}~ \frac{ 2 D ~N^2 N_{\rm t}}{(\Delta x)^2 \Delta t} \delta_{\Omega \Omega'} \delta_{{\bf k} {\bf k}'}.
\end{eqnarray}

The averaged power spectrum of eddies ${\bf v}'$ is (for $k > 0$)
\begin{eqnarray}
\overline{ E({\bf k}, t) } &\equiv& \frac{1}{N^4} \overline{ {\bf v} ({\bf k}, t) \cdot {\bf v}^{*} ({\bf k}, t) } \nonumber\\
&=& \frac{1}{N^4} \frac{1}{N_{\rm t}^2} \sum\limits_{\Omega, \Omega'} \overline{ {\bf v}({\bf k}, \Omega)\cdot {\bf v}^{*}({\bf k}, \Omega') } ~\rme^{-\rmi  (\Omega - \Omega') t } \nonumber\\
&=& \frac{2D}{L^2 T} \left[ \sum\limits_{\Omega }   \frac{1}{(a + \mu k^2)^2 + \Omega^2} \right].
\end{eqnarray}
For long time series (both $T$ and $N_{\rm t}$ are large), the sum can be approximated as an integral
\begin{eqnarray}
 \frac{2 \pi}{T}  \sum\limits_{\Omega}  \frac{1}{(a + \mu k^2)^2 + \Omega^2} &\approx& \int_{-\infty}^{\infty} \rmd \Omega \frac{1}{(a + \mu k^2)^2 + \Omega^2} \nonumber\\
 &=& \frac{\pi}{a + \mu k^2}. \label{eq:freqint}
\end{eqnarray}
Thus the power spectrum
\begin{eqnarray}
\overline{ E({\bf k}, t) } =   \frac{1}{a + \mu k^2}  \frac{  D  }{L^2 } ,
\end{eqnarray}
for $k>0$. 

The isotropic nature of the statistics, specifically $\overline{\psi({\bf k}, \Omega)~ \psi^{*} ({\bf k}, \Omega') }$ being independent of the direction of ${\bf k}$, leads to a useful identity for the velocity correlation. 
The off-diagonal elements of the velocity correlation
\begin{eqnarray}
 \overline{ \langle v_{\rm x}^{'}({\bf r}, t)  v_{\rm y}^{'}({\bf r}, t) \rangle_{\bf r} } &=& \overline{\langle - \partial_y \psi({\bf r}, t) \cdot \partial_x \psi({\bf r}, t) \rangle_{\bf r}}\nonumber\\
 &=&   \frac{2D }{L^2 T}  \sum\limits_{\Omega}  ~  {\sideset{}{'}\sum\limits_{\bf k}}    \frac{-k_{\rm x} k_{\rm y}}{k^2}\frac{1}{(a + \mu k^2)^2 + \Omega^2}  \nonumber \\
 &\propto&  \sum\limits_{n_{\rm x} = -N/2}^{N/2 -1}  \frac{k_{\rm x}}{k^2}\frac{1}{(a + \mu k^2)^2 + \Omega^2} \approx 0.
 \end{eqnarray}
 The diagonal elements
 \begin{eqnarray}
\overline{  \langle  {v_{\rm x}}^{'}({\bf{r}}, t)~ v_{\rm x}^{'} ({\bf{r}}, t)\rangle_{\bf r} } &=&\overline{\langle  (\partial_y \psi)^2  \rangle_{\bf r}} =   \frac{2D }{L^2 T}  \sum\limits_{\Omega}   ~{\sideset{}{'}\sum\limits_{\bf k}}    \frac{k_{\rm y}^2}{k^2}\frac{1}{(a + \mu k^2)^2 + \Omega^2}  \nonumber\\
&=&  \frac{2D }{L^2 T}  \sum\limits_{\Omega}   ~{\sideset{}{'}\sum\limits_{\bf k}}    \frac{k_{\rm x}^2}{k^2}\frac{1}{(a + \mu k^2)^2 + \Omega^2}  \nonumber\\
&=&\overline{\langle ( \partial_x \psi)^2  \rangle_{\bf r}} = \overline{  \langle  {v_{\rm y}^{'}}({\bf{r}}, t)~ v_{\rm y}^{'} ({\bf{r}}, t)\rangle_{\bf r} } \nonumber.
 \end{eqnarray} 
Therefore the velocity correlation
\begin{eqnarray}
\overline{  \langle  {v_i}^{'}({\bf{r}}, t)~ v_j^{'} ({\bf{r}}, t)\rangle_{\bf r} } &=& \frac{1}{2}   \overline{  \langle |\nabla \psi|^2\rangle_{\bf r} } ~\delta_{ij},
\end{eqnarray}
for $i, j = x, y$.

The term 
\begin{eqnarray}
\overline{\langle |\nabla \psi|^2 \rangle_{\bf r} }  &=& \frac{1}{N^6 N_{\rm t}^2} \sum\limits_{\bf r}  ~\sum\limits_{\Omega, \Omega'}~ {\sideset{}{'}\sum\limits_{{\bf k}, {\bf k}'}} ~    ( {\bf k}  \cdot {\bf k'} ) \overline{{\psi}({\bf k}, \Omega) {\psi}^{*}({\bf k'}, \Omega')}  \rme^{- \rmi (\Omega - \Omega')t + i ({\bf k} - {\bf k}')\cdot {\bf r}}  \nonumber\\
   &=&    \frac{2D}{L^2  T}   \sum\limits_{\Omega}   ~ {\sideset{}{'}\sum\limits_{\bf k}}    \frac{1}{(a + \mu k^2)^2 + \Omega^2}. 
 \end{eqnarray}
Using (\ref{eq:freqint}),
\begin{eqnarray}
  \overline{  \langle |\nabla \psi|^2 \rangle_{\bf r}  }&\approx&  \frac{D}{L^2 }   {\sideset{}{'}\sum\limits_{n_{\rm x}, n_{\rm y} = -N/2}^{N/2 -1} }      \frac{1}{a + \mu k^2}. 
  \label{eq: vvcor}
\end{eqnarray} 
For large $L$,  the spacing between $\{ k_i = (2 \pi/L) n_i, ~ i = x, y\}$ becomes very dense, and the above sum can be approximated as an integral. However, in this paper we compare the theoretical prediction with numerical simulation at {\it small} $L$, so the discrete sum is evaluated numerically. 

Since the noise ${\bf f}({\bf r}, t)$ follows a Gaussian distribution with zero mean, any third-order moment $\overline{f_\alpha ({\bf r}_1, t_1)~ f_\beta ({\bf r}_2, t_2)~f_\gamma ({\bf r}_3, t_3) } $ vanishes. It is straightforward to show that this leads to $\overline{ \langle |\nabla \psi|^2 {\bf v}' \rangle_{\bf r} } = 0$.

\section{Finite-size scaling forms in the weak-coupling critical theory \label{app:scaling}}
We derive the exact FSS forms (\ref{eq:cscalingformWCCT})--(\ref{eq:GscalingformWCCT}) from the Fokker-Planck equation (\ref{eq:FP}) of the WCCT. 

The steady-state probability distribution function is the Boltzmann distribution 
\begin{eqnarray}
P_{\rm s.s.}({\bf c}) = \frac{1}{Z} \rme^{-L^2 U/D},
\end{eqnarray}
where the partition function
\begin{eqnarray}
Z = \int \rmd^2 {\bf c} ~ \rme^{-L^2 U/D} = \int \rmd^2 {\bf c} \exp \left[- \frac{a_{\rm R} L^2}{2D} c^2 - \frac{bL^2}{4D} c^4 \right].
\end{eqnarray}

The mean speed
\begin{eqnarray}
\bar{c} &=& \frac{\int \rmd^2 {\bf c}~ c ~\exp[- \frac{a_{\rm R} L^2}{2D} c^2 - \frac{bL^2}{4D} c^4 ]}{\int \rmd^2 {\bf c}~ \exp[- \frac{a_{\rm R} L^2}{2D} c^2 - \frac{bL^2}{4D} c^4]}  \nonumber\\
&=& \frac{\int_0^\infty \rmd c~ c^2 ~\exp[- \frac{a_{\rm R} L^2}{2D} c^2 - \frac{bL^2}{4D} c^4 ]}{\int_0^\infty \rmd c~c~ \exp[- \frac{a_{\rm R} L^2}{2D} c^2 - \frac{bL^2}{4D} c^4 ]}.  
\end{eqnarray}
Introducing a change of variable from $c$ to $\tilde{c} = c \left( \frac{bL^2}{4D} \right)^{1/4}$, we obtain
\begin{eqnarray}
\bar{c} & =& \left( \frac{4D}{bL^2} \right)^{1/4}   \frac{ \int_0^\infty  \rmd \tilde{c} ~\tilde{ c}^2 ~ \exp [ - \frac{a_{\rm R} L}{\sqrt{bD}} \tilde{c}^2 -  \tilde{c}^4 ]}{\int_0^\infty \rmd \tilde{c}~\tilde{c}~ \exp [ - \frac{a_{\rm R} L}{\sqrt{bD}} \tilde{c}^2 -  \tilde{c}^4 ]}\nonumber\\
&=& L^{-1/2} \cdot F_1 (a_{\rm R}  L) = L^{-1/2} \cdot F_1 [(a - a_0)  L] ,
\end{eqnarray}
where the scaling function 
\begin{eqnarray}
F_1(x) &\equiv& \left( \frac{4D}{b} \right)^{1/4}   \frac{g_2(x; b, D) }{ g_1(x; b, D)}, 
\end{eqnarray}
and the integral $g_n(x; b, D)$ is defined as
\begin{eqnarray}
g_n (x; b, D) \equiv  \int_0^\infty  d {c} ~{ c}^n ~ \exp \left[ - \frac{x}{\sqrt{bD}} {c}^2 -  {c}^4 \right].
\end{eqnarray}
Likewise the second-order moment
\begin{eqnarray}
\bar{c^2} &=& \left( \frac{4D}{bL^2} \right)^{1/2}  \frac{g_3 (a_{\rm R}L; b, D) }{g_1 (a_{\rm R}L; b, D) },
\end{eqnarray}
and the susceptibility 
\begin{eqnarray}
\chi &\equiv& L^2 [\bar{c^2} - \bar{c}^2] = L \cdot  F_2 [a_{\rm R}  L]= L \cdot  F_2 [(a - a_0)  L], 
\end{eqnarray}
where the scaling function 
\begin{eqnarray}
F_2(x) &\equiv&  2 \sqrt {\frac{D}{b}}  \frac{g_3(x; b, D)}{g_1(x; b, D)} - F_1^2 (x). \label{eq:F2scalingfunc}
\end{eqnarray}
The fourth-order moment
\begin{eqnarray}
\bar{c^4} &=&  \left( \frac{4D}{bL^2} \right)  \frac{g_5 (a_{\rm R}L; b, D) }{g_1 (a_{\rm R}L; b, D) },
\end{eqnarray}
and the Binder cumulant 
\begin{eqnarray}
G &\equiv& 1 - \frac{\bar{c^4}}{3 (\bar{c^2})^2} = F_3(a_{\rm R} L) = F_3[(a - a_0) L],
\end{eqnarray}
where the scaling function
\begin{eqnarray}
F_3(x) \equiv  1 -  \frac{1}{3} \frac{g_5 (x; b, D) \cdot g_1 (x; b, D) }{g_3^2 (x; b, D) } .
\end{eqnarray}

\bibliographystyle{iopart-num}
\bibliography{ref}

\end{document}